\documentclass[aps,prd,reprint,preprintnumbers,floats,epsfig,nofootinbib,amssymb,nofootinbib]{revtex4}

\usepackage{slashed}
\usepackage{graphicx,color}
\usepackage{epsfig}
\usepackage{subfigure}
\usepackage{epsfig}
\usepackage{epstopdf}
\usepackage{dcolumn}
\usepackage{bm}
\usepackage{color}
\usepackage{ulem}
\usepackage{enumitem}
\usepackage{multirow}
\usepackage[colorlinks,
            linkcolor=black,
            anchorcolor=black,
            citecolor=black
            ]{hyperref}

\begin{document}

\baselineskip=15pt
%\preprint{}

\title{Electroweak precision tests for triplet scalars}

\author{Yu Cheng$^{1}$\footnote{chengyu@sjtu.edu.cn}}
\author{Xiao-Gang He${}^{1,2}$\footnote{hexg@sjtu.edu.cn}}
\author{Fei Huang${}^{1}$\footnote{fhuang@sjtu.edu.cn}}
\author{Jin Sun$^{1}$\footnote{019072910096@sjtu.edu.cn}}
\author{Zhi-Peng Xing$^{1}$\footnote{zpxing@sjtu.edu.cn}}

\affiliation{${}^{1}$Tsung-Dao Lee Institute,  KLPAC  and SKLPPC, and School of Physics and Astronomy, Shanghai Jiao Tong University, Shanghai 200240, China}
\affiliation{${}^{2}$National Center for Theoretical Sciences, and Department of Physics, National Taiwan University, Taipei 10617, Taiwan}

\begin{abstract}
Electroweak precision observables are fundamentally important for testing the standard model (SM) or its extensions. The influences to observables from new physics  within the electroweak sector can be  expressed in terms of oblique parameters S, T, U.  The  recently reported $W$ mass excess anomaly by CDF modifies these parameters in a significant way.  By performing the global fit with the CDF new $W$ mass  measurement data, we obtain $S=0.03 \pm 0.03$, $T=0.06 \pm 0.02$ and $U=0.16 \pm 0.03$ (or $S=0.14 \pm 0.03$, $T=0.24 \pm 0.02$ with $U=0$) which is significantly away from zero as SM would predict.
The CDF excess  strongly indicates the need of  new physics beyond SM. 
We carry out  global fits to study the influence of two different cases of simple extensions by  a hyper-charge $Y=1$ triplet scalar $\Delta$ (corresponding to the type-II seesaw model) and a $Y=0$  real triplet scalar $\Sigma$, on electroweak precision tests  to determine parameter space in these models to solve the $W$ mass anomaly and discuss the implications.
We find that these triplets can affect the oblique parameters significantly at the tree and loop levels.
For $Y=1$ case, there are seven new scalars in the model. 
The tree and scalar loop effects on oblique parameters can be expressed in terms of three parameters, the doubly-charged   mass $m_{H^{++}}$, potential parameter $\lambda_4$  and triplet vev $v_\Delta$. 
Our global fit  obtains $m_{H^{++}}=103.02 \pm 9.84$ GeV, $\lambda_4=1.16 \pm 0.07$  and $v_\Delta=0.09 \pm 0.09$ GeV. 
These parameter correspondingly results in mass difference $\Delta m=m_{H^+}-m_{H^{++}}=64.78\pm 2.39$ GeV.
 We find that the doubly-charged mass $m_{H^{++}}$ satisfies the current LHC constraints. We also further adopt the phenomenological analysis, such as vacuum stability and perturbative unitarity, Higgs data, triple Higgs self-coupling and lepton colliders analysis. 
 %Furthermore, we find that the future lepton colliders can feasibly search for doubly charged $H^{++}$    via pair production $H^{++}H^{--}$ and decay  channel to four $W$ final states.
For $Y=0$ case, there are four new scalars in the model.
 The tree and loop effects on the oblique parameters can be parameterized  by the singly charged mass $m_{H^+}$, the mass difference $\Delta m=m_{H^+}-m_{H^0}$ and triplet vev $v_\Sigma$ with the fit values $m_{H^+}=199.45 \pm 39.73$ GeV, $\Delta m=-2.32\pm1.99$ GeV and $v_\Sigma=3.86\pm 0.27$ GeV, respectively.
% Our global fit obtains  $m_{H^+}=199.45 \pm 39.73$ GeV, $\Delta m=-2.32\pm1.99$ GeV and $v_\Sigma=3.86\pm 0.27$ GeV. 
However, these strongly violate the perturbative unitarity of the  potential parameter $b_4$, which can  be satisfied within $1\sigma$ errors. 
\end{abstract}

\maketitle

\section{Introduction}

Recently CDF collaboration announced their new measurement of $W$ boson mass with a value of~\cite{CDF} $80,433.5 \pm 9.4 $ MeV which is 7$\sigma$ above the standard model (SM) prediction~\cite{pdg} of $80,357 \pm 6 $ MeV. This is a significant indication of new physics beyond the SM (BSM). A lot of efforts have been made to provide an explanation for this tantalizing excess,
such as singlet scalar model~\cite{Sakurai:2022hwh,Peli:2022ybi,Dcruz:2022dao,Asai:2022uix}, Two-Higgs Doublet model~\cite{Fan:2022dck,Lu:2022bgw,Song:2022xts,Bahl:2022xzi,Babu:2022pdn,Heo:2022dey,Ahn:2022xeq,Ghorbani:2022vtv,Lee:2022gyf,Abouabid:2022lpg,Benbrik:2022dja,Botella:2022rte,Kim:2022hvh,Kim:2022xuo,Appelquist:2022qgl,Benincasa:2022elt,Arhrib:2022inj,Han:2022juu,Abdallah:2022shy}, 
triplet scalar model~\cite{Barrie:2022cub,Cheng:2022jyi,Du:2022brr,FileviezPerez:2022lxp,Kanemura:2022ahw,Mondal:2022xdy,Borah:2022obi,Addazi:2022fbj,Heeck:2022fvl,Chen:2022ocr,Evans:2022dgq,Ghosh:2022zqs,Ma:2022emu,Bahl:2022gqg,Penedo:2022gej}, 
new fermion~\cite{Blennow:2022yfm,Lee:2022nqz,Cheung:2022zsb,Crivellin:2022fdf,Ghoshal:2022vzo,Kawamura:2022uft,Popov:2022ldh,Cao:2022mif,Dermisek:2022xal,Li:2022gwc,He:2022zjz,Chowdhury:2022dps}, 
new gauge boson~\cite{Zeng:2022llh,Zhang:2022nnh,Zeng:2022lkk,Du:2022fqv,Baek:2022agi,Cheng:2022aau,Faraggi:2022emm,Cai:2022cti,Thomas:2022gib,Wojcik:2022rtk,Afonin:2022cbi,Allanach:2022bik,Nagao:2022dgl,VanLoi:2022eir}, 
effective field theory~\cite{deBlas:2022hdk,Fan:2022yly,Bagnaschi:2022whn,Paul:2022dds,Balkin:2022glu,Cirigliano:2022qdm,Almeida:2022lcs,Gupta:2022lrt,Guedes:2022cfy,Liu:2022vgo}, 
SUSY~\cite{Yang:2022gvz,Du:2022pbp,Tang:2022pxh,Athron:2022isz,Sun:2022zbq,Zheng:2022irz}  and axion~\cite{Lazarides:2022spe}, and the different combinations of the above new particles~\cite{Yuan:2022cpw,Coy:2021hyr,DAlise:2022ypp,Strumia:2022qkt,Arias-Aragon:2022ats,Asadi:2022xiy,DiLuzio:2022xns,Gu:2022htv,Biekotter:2022abc,DiLuzio:2022ziu,Endo:2022kiw,Nagao:2022oin,Arcadi:2022dmt,Chowdhury:2022moc,Bhaskar:2022vgk,Borah:2022zim,Batra:2022org,Batra:2022pej,Wang:2022dte,Li:2022eby,Kawamura:2022fhm,Zhou:2022cql,Rizzo:2022jti}.

The $W$ mass measurement with high precision is very important to test the SM and also new physics (NP) models beyond. It can affect many other observables in a given model. One effective way commonly used to study its effects is to take known relevant data together to carry out a global fit, known as the electroweak (EW) fit. This method involves fitting over a set of well-measured SM observables, and minimizing the $\chi^2$ value over both the fitted observables. The EW fit leverages the small uncertainties of the fitted observables to probe precision predictions of the derived observables for a given theoretical model. Thus, 
the global EW fit is a powerful tool to explore the correlations among observables in the SM and predict the new physics. Due to CDF new excess in $W$ mass measurement, we can expect some observables in the global fits may suffer from new tension since the EW parameters in the SM are closely related to each other.
An impressive amount of activity has already been generated by  CDF experiment to perform global EW fit with the new input  $m_W$~\cite{deBlas:2022hdk,Fan:2022yly,Asadi:2022xiy,Gu:2022htv,Endo:2022kiw,Liu:2022vgo,Lu:2022bgw,Miralles:2022jnv,Carpenter:2022oyg,Almeida:2022lcs,Balkin:2022glu,Strumia:2022qkt}. These fits are mainly performed by the oblique  parameters $S$, $T$ and $U$~\cite{Peskin:1990zt,Peskin:1991sw} or Wilson coefficients in EFT. However, the correlations between $S$, $T$ and $U$ depend on different models, a global fit stopping at these oblique parameters may not provide intricate detail about the models.  Therefore,
%We find that most of the recent global studies are in lacking to perform the fit via directly NP parameters.
it is imperative to analyze the  NP effects by adopting directly the new parameters to perform fits. 

When going beyond SM, there are many different ways where the electroweak oblique parameters can be affected.  In this work, we study the impact of $SU(2)_L$ triplet scalar extensions in SM $SU(3)_C\times SU(2)_L\times U(1)_Y$ gauge group on the oblique parameters in light of the recent CDF data by carrying out  global fits and the related phenomenology analysis. We focus on two kinds of interesting triplet models, hyper-charge $Y=1$ triplet $\Delta: (1,3)(1)$ and $Y=0$ triplet $\Sigma: (1,3)(0)$ models. Here the numbers in the first bracket indicate the $SU(3)_C$ and $SU(2)_L$ quantum numbers, and the ones in the second bracket indicate the $U(1)_Y$ quantum numbers.
The $Y=1$ case is the famous type-II seesaw model which is well motivated by providing small neutrino mass~\cite{Magg:1980ut,Cheng:1980qt,Lazarides:1980nt,Mohapatra:1980yp}.  The $Y=0$ case also has many interesting features for dark matter and dark photon model buildings~\cite{Ross:1975fq,Gunion:1989ci,Forshaw:2001xq,Forshaw:2003kh,Chen:2006pb,Chankowski:2006hs,Chardonnet:1993wd,Blank:1997qa,Chivukula:2007koj,  Cirelli:2005uq,Cirelli:2007xd,Fuyuto:2019vfe,Cheng:2021qbl}. 
In principle, the electric neutral components can develop vacuum expectation values (vev) inducing  $T$ parameter deviation from the SM prediction at the tree level. The component triplet scalars can have different masses from Higgs potential which then induce $S$, $T$ and $U$ deviations  from SM predictions at loop level. In the global fit we use the new scalar parameters to present the influences on the oblique parameters including the tree-level and one loop level contributions simultaneously. We  further adopt our fit results to perform  the relevant phenomenological analysis. 

This paper is organized as follows. In Sec. II, we show the details of the Higgs triplet models. Sec. III performs the EW global fits.  Sec. IV analyzes the relevant phenomenology.   In Sec. V, we  draw our conclusion.

\section{The $Y=1$ and $Y=0$ Higgs triplet models}

In this section we give some details about the two different cases of Higgs triplet models with $Y=1$ and $Y=0$.

\subsection{$Y=1$ triplet scalar}

Extend the scalar sector with a $Y=1$ triplet $\Delta$ in addition to the usual SM doublet scalar $\phi: (1,2)(1/2)$ will modify the SM Lagrangian in two ways, making the Higgs potential $V(\phi, \Delta)$ to depend on  two scalars and introducing a new Yukawa coupling term 
$L_Y = y_{ij} \bar L^c_{iL} i \tau_2 \Delta L_{jL}$. 
The  coupling will induce non-zero neutrino masses which is the core of type-II seesaw model. And the effects on the oblique parameters come from the two parts, triplet  scalar vev and mass splitting of the component fields. 
We will  study the Higgs potential and its implication on the oblique parameters in the following.
% and further analyze phenomenology of the Yukawa coupling term after our global fit to the oblique parameters.

The component fields  of $\Phi$ and $\Delta$ are given by
\begin{eqnarray}
		\phi = \left (\begin{array}{c}   \phi^+ \\  \; \frac{1}{\sqrt{2}}(v+\phi^0+i\chi) \end{array} \right )\;, \;\;\;
	\Delta = \left (\begin{array}{cc}   \Delta^+/\sqrt{2}&\;  \Delta^{++}\\  \Delta^0&\; - \Delta^+/\sqrt{2} \end{array} \right )\;,\;\;\;\Delta^0=\frac{v_\Delta+\delta+i\eta}{\sqrt{2}}\;. 
\end{eqnarray}
Here $v^2+2v_\Delta^2=v^2_{SM}=(246\mbox{GeV})^2$. 

The tree and loop level will contribute to modify the oblique parameters at the same time.  The tree level one is from the vev $v_\Delta$ modifications to W and Z masses with $m^2_W=g^2 v^2_{SM}/4$ and $m^2_Z=g^2\left(v^2+4v_\Delta^2\right)/4c_W^2$, which leads to the EW $\rho$ parameter defined by $\rho=m_W^2/(m_Z^2 c_W^2)$ as~\cite{Ross:1975fq} 
\begin{eqnarray}\label{tree}
	\rho= 1 - \frac{2v_\Delta^2}{(v^2+4v_\Delta^2)}\;.
\end{eqnarray}
This is related to the $T$ parameter by $T_{tree} = -2 v^2_\Delta/(v^2+4v^2_\Delta)/\alpha_{em}$. Here $\alpha_{em}$ is the electromagnetic fine structure constant.   

As the $\rho$ parameter is constrained to be close to 1~\cite{pdg}, we can obtain qualitatively $v_\Delta< 5$ GeV within $3\sigma$. This indicates that $v_\Delta<<v$ is a good approximation, which will be used in  the following discussion. Since in this model the neutrino masses are proportional to $v_\Delta$, a small $v_\Delta$ helps to explain why neutrino masses are much smaller than their charged lepton partners. Note that the increased $W$ mass from CDF makes the $\rho$ parameter larger than previous one, therefore a non-zero $v_\Delta$ tends to worsen the $W$ mass excess problem. 
However, there are additional one loop contributions to the oblique parameters to solve  the problem. We now study these contributions.

The most general scalar potential with doublet and triplet is given by
\begin{eqnarray}
	V(\phi,\Delta) &=& -m_\phi^2 (\phi^+ \phi )+\frac{\lambda}{4}(\phi^+ \phi )^2 
	+\lambda_1 (\phi^+ \phi) Tr (\Delta^+ \Delta) +\lambda_2 (Tr (\Delta^+ \Delta))^2\nonumber\\
	&+&\lambda_3 Tr (\Delta^+ \Delta \Delta^+ \Delta)
	+\lambda_4 \phi^+ \Delta \Delta^+ \phi 
	+\tilde M_\Delta^2 Tr (\Delta^+ \Delta) +(\mu \phi^T i \tau_2 \Delta^+ \phi +h.c.)\;. 
\end{eqnarray}
%{\color{red}Here $\mu$ term will source the lepton number violation and denote the soft $Z_2$ breaking  with assuming to be real, %which will affect the mixing between doublet $H$ and triplet $\Delta$. }
After electroweak symmetry breaking, minimization for the potential gives 
\begin{eqnarray}
	\tilde M^2_\Delta &=& \frac{\mu v^2}{\sqrt{2}v_\Delta}-\frac{\lambda_1+\lambda_4}{2}v^2
	-(\lambda_2+\lambda_3)v_\Delta^2\;,\nonumber\\
	m_\phi^2 &=&\frac{\lambda}{4}v^2+ \frac{\lambda_1+\lambda_4}{2}v_\Delta^2-\sqrt{2}\mu v_\Delta\;. 
\end{eqnarray}
Here we can define $M_\Delta^2=\mu v^2/(\sqrt{2}v_\Delta)$. A small $v_\Delta$ implies a large $M_\Delta$.

The mass of the doubly charged  scalar bosons  $H^{\pm\pm} (\Delta^{\pm\pm})$  is
\begin{eqnarray}
	m_{H^{\pm\pm}}^2 = \frac{(\sqrt{2}\mu v^2-\lambda_4 v^2 v_\Delta-2\lambda_3 v_\Delta^3)}{2v_\Delta} \;. 
\end{eqnarray}

Mass eigenstates of the singly-charged states, CP-odd neutral states and CP-even neutral states
are obtained by the corresponding mixing angles $\beta_+$, $\beta$ and $\alpha$
\begin{eqnarray}
		\left (\begin{array}{l}
		G^+\\
		H^+
	\end{array}
	\right ) = \left (\begin{array}{cc}  \cos\beta_+\;\;\;&\; \;\; \sin\beta_+\\  -\sin \beta_+\;\;\;&\; \;\; \cos\beta_+ \end{array} \right )
	\left (\begin{array}{l}
		\phi^+\\
		\Delta^+
	\end{array}
	\right )\;,\;\;
		\left (\begin{array}{l}
		G\\
		A
	\end{array}
	\right ) = \left (\begin{array}{cc}  \cos\beta\;\;\;&\; \;\; \sin\beta\\  -\sin \beta\;\;\;&\; \;\; \cos\beta \end{array} \right )
	\left (\begin{array}{l}
		\chi\\
		\eta
	\end{array}
	\right )\;,\;\;
	\left (\begin{array}{l}
		h\\
		H
	\end{array}
	\right ) = \left (\begin{array}{cc}  \cos\alpha\;\;\;&\; \;\; \sin\alpha\\  -\sin \alpha\;\;\;&\; \;\; \cos\alpha \end{array} \right )
	\left (\begin{array}{l}
		\phi^0\\
		\delta
	\end{array}
	\right )\;,
\end{eqnarray}
where $G^+(G)$  are Nambu-Goldstone bosons which will be eaten by $W^+(Z)$  to provide the longitudinal components. Note that the physical fields are singly charged $H^+$, neutral CP-odd $A$, SM-like $h$ and neutral CP-even $H$.

The mass squared matrix for singly charged  field  is
\begin{eqnarray}
	M_{\pm}^2 = \left(\sqrt{2}\mu-\frac{\lambda_4 v_\Delta}{2}\right)\left (\begin{array}{cc}   v_\Delta\;\;\;&\; \;\; -\frac{v}{\sqrt{2}}\\ \\
		-\frac{v}{\sqrt{2}}\;\;\;&\; \;\; \frac{ v^2}{2v_\Delta} \end{array} \right )\;,
\end{eqnarray}
By diagonalizing the mass matrix with the mixing angle $\beta_+$, we obtain 
the $H^\pm$ mass as
\begin{eqnarray}
	m_{H^\pm}^2 =(2\sqrt{2}\mu-\lambda_4 v_\Delta) \frac{(2v_\Delta^2+ v^2)}{4v_\Delta} \;,\;\;\;
	\mbox{with}\;\;\cos\beta_+=\frac{v}{\sqrt{v^2+2v_\Delta^2}}\;.
\end{eqnarray}

The mass squared matrix for neutral CP-odd  field is
\begin{eqnarray}
	M_{odd}^2 = \sqrt{2}\mu\left (\begin{array}{cc}   2v_\Delta\;\;\;&\; \;\; -v\\  -v\;\;\;&\; \;\; \frac{ v^2}{2v_\Delta} \end{array} \right )\;. 
\end{eqnarray}
By diagonalizing the mass matrix with the mixing angle $\beta$, we obtain 
the $A$ mass as
\begin{eqnarray}
	m_A^2 =\mu \frac{(4v_\Delta^2+ v^2)}{\sqrt{2}v_\Delta} \;.\;\;\;
	\mbox{with}\;\; \cos\beta=\frac{v}{\sqrt{v^2+4v_\Delta^2}}\;.
\end{eqnarray}

The mass squared matrix for neutral CP-even  fields  are
\begin{eqnarray}
	M_{even}^2 = \left (\begin{array}{cc}   \frac{\lambda}{2}v^2\;\;\;&\; \;\; (\lambda_1+\lambda_4)vv_\Delta-\sqrt{2}\mu v\\  (\lambda_1+\lambda_4)vv_\Delta-\sqrt{2}\mu v\;\;\;&\; \;\; \frac{\mu v^2}{\sqrt{2}v_\Delta}+2(\lambda_2+\lambda_3)v_\Delta^2 \end{array} \right )\;. 
\end{eqnarray}
By diagonalizing the mass matrix with the mixing angle $\alpha$, we obtain the masses of $h$ and $H$ as 
\begin{eqnarray}
	&&m_h^2 = \frac{1}{2}\left[\frac{\lambda}{2}v^2+ \frac{\mu v^2}{\sqrt{2}v_\Delta}+2(\lambda_2+\lambda_3)v_\Delta^2-\left|\left(\frac{\lambda}{2}v^2-\frac{\mu v^2}{\sqrt{2}v_\Delta}-2(\lambda_2+\lambda_3)v_\Delta^2\right)\right|\sqrt{1 +\tan^2 2\alpha}  \right]\;,\nonumber\\
	&&m_H^2 = \frac{1}{2}\left[\frac{\lambda}{2}v^2+ \frac{\mu v^2}{\sqrt{2}v_\Delta}+2(\lambda_2+\lambda_3)v_\Delta^2+\left|\left(\frac{\lambda}{2}v^2-\frac{\mu v^2}{\sqrt{2}v_\Delta}-2(\lambda_2+\lambda_3)v_\Delta^2\right)\right|\sqrt{1 +\tan^2 2\alpha}  \right]\;,\nonumber\\
	&&\mbox{with}\;\;	\tan 2\alpha= \frac{2[(\lambda_1+\lambda_4)vv_\Delta-\sqrt{2}\mu v]} { \frac{\lambda}{2}v^2-\frac{\mu v^2}{\sqrt{2}v_\Delta}-2(\lambda_2+\lambda_3)v_\Delta^2}
\end{eqnarray}
Here we identify $h$ as the SM-like Higgs with $m_h=125$ GeV. 
Note that for $\lambda_i \sim O(1)$, $\alpha$ will be small due to the suppression of $v_\Delta/v$. Therefore,  $\alpha\approx 0$ is a good approximation. In this case,   the couplings with the SM Higgs boson will not be modified much compared with SM ones.

In the limit $v_\Delta << v$, the above physical mass splittings  are \cite{Mandal:2022zmy}
\begin{eqnarray}\label{mass splittings}
	m_\Delta^2 \approx m_{H^0}^2 \approx m_{A^0}^2 \approx m_{H^+}^2+\frac{\lambda_4}{4}v^2
	\approx m_{H^{++}}^2+\frac{\lambda_4}{2}v^2\;.
\end{eqnarray}

To compare with EW precision experimental data, we now recast the triplet effects in terms of oblique parameters $S$, $T$ and $U$ via modifying the EW gauge boson self-energies~\cite{Peskin:1990zt,Peskin:1991sw}. 
The tree level case will only affect the $T$ parameter as shown in Eq.~(\ref{tree}).

At one loop level, we obtain the oblique parameters   by choosing the weak isospin $J=1$ and hypercharge $Y=1$ in Ref.~\cite{Lavoura:1993nq}
\begin{eqnarray}\label{loop}
	S&=&-\frac{1}{3\pi}  \ln \frac{m_{H^{++}}^2}{m_{H^0}^2}
	-\frac{2}{\pi}\left[(1-2 s_W^2)^2\xi\left(\frac{m^2_{H^{++}}}{m^2_{Z}},\frac{m^2_{H^{++}}}{m^2_{Z}}\right)
	+ s_W^4\xi\left(\frac{m^2_{H^+}}{m^2_{Z}},\frac{m^2_{H^+}}{m^2_{Z}}\right)	
	+\xi\left(\frac{m^2_{H^0}}{m^2_{Z}},\frac{m^2_{H^0}}{m^2_{Z}}\right)
	\right]\;,\nonumber\\
	T&=&\frac{1}{8\pi c_W^2 s_W^2} \left[ \eta\left(\frac{m^2_{H^{++}}}{m_Z^2},\frac{m^2_{H^+}}{m_Z^2}\right)
	+\eta\left(\frac{m^2_{H^{+}}}{m_Z^2},\frac{m^2_{H^0}}{m_Z^2}\right)\right]\;,\nonumber\\
	U&=&\frac{1}{6\pi } \ln \frac{m_{H^+}^4}{m_{H^{++}}^2 m_{H^0}^2}
	+\frac{2}{\pi} 
	\left[ (1-2 s_W^2)^2\xi\left(\frac{m^2_{H^{++}}}{m_Z^2},\frac{m^2_{H^{++}}}{m_Z^2}\right)
	+	s_W^4\xi\left(\frac{m^2_{H^{+}}}{m_Z^2},\frac{m^2_{H^{+}}}{m_Z^2}\right)
	+\xi \left(\frac{m^2_{H^0}}{m_Z^2},\frac{m^2_{H^0}}{m_Z^2}\right)\right]\nonumber\\
	&-&\frac{2}{\pi} 
	\left[ 
	\xi\left(\frac{m^2_{H^{++}}}{m_W^2},\frac{m^2_{H^+}}{m_W^2}\right)
	+\xi\left(\frac{m^2_{H^{+}}}{m_W^2},\frac{m^2_{H^0}}{m_W^2}\right)\right]\;.
\end{eqnarray}
where the functions are defined as
\begin{eqnarray}
	&&\xi(x,y)=\frac{4}{9}-\frac{5}{12} (x+y)+\frac{1}{6}(x-y)^2+\frac{1}{4}\left[x^2-y^2-\frac{1}{3}(x-y)^3
	-\frac{x^2+y^2} {x-y}  \right] \ln \frac{x}{y}-\frac{1}{12}d(x,y)f(x,y)\;,\nonumber\\
	&&\eta(x,y)=x+y-\frac{2xy}{x-y} \ln \frac{x}{y}\;,\nonumber\\
	&&d(x,y)=-1+2(x+y)-(x-y)^2\;\nonumber\\
	&&
	f(x,y)=\left\{
	\begin{array}{cl}
		-2\sqrt{d(x,y)}\left[\arctan \frac{x-y+1}{\sqrt{d(x,y)}}
		-\arctan \frac{x-y-1}{\sqrt{d(x,y)}}\right]\;,\;\; &  \;\;\; d(x,y) > 0 \\
		\sqrt{-d(x,y)}\ln \frac{x+y-1+\sqrt{-d(x,y)}}{x+y-1-\sqrt{-d(x,y)}}\;,\;\; &  \;\;\; d(x,y) \le 0
	\end{array} \right.
\end{eqnarray}
We find that the above functions are  symmetric under the exchange of $x$ and $y$. If one only focuses on heavy new particles and small mass splitting, $|y-x|/x << 1 <<4x-1$, the above functions can be further approximated to the leading order as
\begin{eqnarray}
	\eta(x,y) =\frac{(x-y)^2}{3x}\;,\;\;\;\;\;\;\;\;
	\xi(x,y)=\frac{1}{60x}\;.
\end{eqnarray}

Our results are consistent with Refs.~\cite{Mandal:2022zmy, Chun:2012jw} and we point out 
 one typo for  U term in Ref.~\cite{Chun:2012jw}. Using the mass splittings in Eq.~(\ref{mass splittings}), 
 the above oblique parameters are further approximated to the leading order as
\begin{eqnarray}\label{loopsimple}
	S&=&\frac{\lambda_4}{6\pi}   \frac{v^2}{m_{H}^2}
	-\frac{m_Z^2}{30\pi m_{H}^2}(2-4s_W^2+5s_W^4)\;,\nonumber\\
	T&=&\frac{1}{12\pi^2 \alpha_{em} }\left( \frac{\lambda_4^2 v^2}{16m_{H}^2}+4m^2_{H}\frac{v^4_\Delta}{v^6}\right)\;,\nonumber\\
	U&=&\frac{1}{6\pi}\frac{4v_\Delta^2}{v^2}+\frac{m_Z^2}{30\pi m^2_{H} } (-2s_W^2+5s_W^4)\;.
\end{eqnarray}
We have kept terms of $v_\Delta^2/v^2$ which have been neglected in previous studies~\cite{Mandal:2022zmy,Heeck:2022fvl}. 

Previously the above two  effects for $W$-mass have been  investigated individually.   The tree-level case is only considered in  Ref.~\cite{Cheng:2022jyi} and the one loop one is analyzed in Ref.~\cite{Heeck:2022fvl,Bahl:2022gqg,Kanemura:2022ahw}.
% where one-loop dominates over the tree level contribution  so that the tree level  can be neglected.
%However,  for  $v_\Delta$ with the few GeV, the tree level may have the large effect making the analysis only with loop effect  inappropriate.
However, the overall triplet contributions to the oblique  parameters should contain the tree-level one in Eq.~(\ref{tree}) and one-loop in Eq.~(\ref{loop}) simultaneously. Especially, $v_\Delta$ with the few GeV may make the tree level and one loop contribution at the same magnitude. Furthermore, the global fits should be performed by NP parameters directly rather than the oblique parameters indirectly. To ensure the completeness of the fit results, we choose the full formula in Eq.~(\ref{loop}) rather than the approximate one in Eq.~(\ref{loopsimple}) to conduct the analysis.

Before performing the EW global  fit, we want to discuss the current experimental constraints on type-II seesaw model. The collider searches constrain  the doubly charged scalar $H^{++}$ mass, which depends on the dominant decay modes are $H^{++}\to l^+l^+$ or $H^{++}\to W^+ W^+$.  If the dominant one is $H^{++}\to l^+l^+$, ATLAS~\cite{ATLAS:2017xqs} (CMS~\cite{CMS:2017pet}) set the lower bounds 770-870 GeV (800-820 GeV)   by searching for doubly charged scalar in the same-charge dilepton invariant mass spectrum. If the dominant one is 
$H^{++}\to W^+W^+$, ATLAS using 8 (13) TeV data excludes the range $m_{H^{++}}<84$ GeV~\cite{Kanemura:2014goa,Kanemura:2014ipa} ($200<m_{H^{++}}<350$ GeV~\cite{ATLAS:2018ceg,ATLAS:2021jol})  by searching for $pp\to H^{++}H^{--} \to 4W$ channel.

Therefore, we need to first analyze the two kinds of decay modes in order to decide the parameter range we choose. The decays into leptons are proportional to the Yukawa coupling while the decays into two W's are proportional to the vev. The relative decay ratios can be estimated  by~\cite{FileviezPerez:2008jbu}
\begin{eqnarray}\label{br}
	\frac{\Gamma(H^{++}\to l^+l^+)}{\Gamma(H^{++}\to W^+W^+)}&\approx&\left(\frac{m_\nu}{m_{H^{++}}}\right)^2   \left(\frac{v}{v_\Delta}\right)^4  
	\;.
\end{eqnarray}
If  $m_\nu/m_{H^{++}}\sim 1 \rm{eV}/ 1 TeV$, one finds that these two decay modes are comparable when   $v_\Delta \sim 10^{-4}$ GeV.
For  $v_\Delta \sim 1$ GeV, one finds that the ratios is around $ 10^{-12}$, which strongly indicates the dominant decay mode is two W's final state. 

Because  the loop contribution  in Eq.~(\ref{loopsimple}) is inversely proportional to new scalar mass, we need large loop contribution to eliminate the negative contribution from the tree level in Eq.~(\ref{tree}). Therefore,  we focus on the range $84<m_{H^{++}}<200$ GeV, which means that the dominant decay mode should be $H^{++}\to W^+W^+$.
\\

\subsection{Triplet scalar with $Y=0$}

Extend the scalar section with a $Y=0$ real triplet $\Sigma$ in addition to the usual SM doublet scalar $\phi: (1,2)(1/2)$ will also modify the SM Lagrangian, making the Higgs potential $V(\phi, \Sigma)$  depend on  two scalars. The modifications to the oblique parameters come from the triplet  vev and mass splitting of the component fields which are determined from Higgs potential.

The component fields  of the triplet $\Sigma$ are given by
\begin{eqnarray}
	 \Sigma =\frac{1}{2} \left (\begin{array}{cc}   \Sigma^0&\; \sqrt{2}\Sigma^+\\ \sqrt{2} \Sigma^-&\; -\Sigma^0 \end{array} \right )\;,\;\;\;\Sigma^0=v_\Sigma+\sigma\;. 
\end{eqnarray}
Here $v^2+4v_\Sigma^2=v^2_{SM}=(246\mbox{GeV})^2$.

A non-zero vev $v_\Sigma$ in this case  will give the gauge boson masses  as $m^2_W=g^2 (v^2+ 4 v^2_\Sigma)/4$ and $m^2_Z=g^2 v^2/4c_W^2$. The EW $\rho$ parameter is given by
\begin{eqnarray}\label{tree1}
	\rho=1+\frac{4v_\Sigma^2}{v^2}\;,
\end{eqnarray}
which leads to an effective $T$ parameter $T_{tree} = 4 v^2_\Sigma/(v^2 \alpha_{em})$. 

Again because the $\rho$ parameter is restricted to be very close to 1~\cite{pdg}, $v_\Sigma$ is constrained to be small. We obtain qualitatively $v_\Sigma< 4$ GeV within $3\sigma$. This also indicates that $v_\Sigma<<v$ is a good approximation. However, unlike the $Y=1$ triplet case, a non-zero $v_\Sigma$ contributes  to the $\rho$ parameter with the right sign and it helps to solve the $W$ mass excess problem.

To obtain the new scalar contributions to the oblique parameters, we carry out a similar analysis as that we did for the $Y=1$ triplet. 
We first analyze the Higgs potential to obtain the mass spectrum of the scalars and then obtain the one loop contribution. 

The most general scalar potential is given by
\begin{eqnarray}
	V(\phi,\Sigma) &=& -m_\phi^2 (\phi^+ \phi )+\lambda_0 (\phi^+ \phi )^2 - M_\Sigma^2 Tr (\Sigma^2)
	+\lambda_1  Tr (\Sigma^4) +\lambda_2 (Tr (\Sigma^2))^2
	\nonumber\\ &+&
	\alpha (\phi^+ \phi) Tr (\Sigma^2)
	+\beta  \phi^+ \Sigma^2 \phi 
	+a_1  \phi^+ \Sigma \phi \;. 
\end{eqnarray}
A more compact form of the above potential can be presented as
\begin{eqnarray}
	V(\phi,\Sigma) &=& -m_\phi^2 (\phi^+ \phi )+\lambda_0 (\phi^+ \phi )^2 - \frac{M_\Sigma^2}{2} F
+\frac{b_4}{4} F^2+a_1  \phi^+ \Sigma \phi + \frac{a_2}{2} F (\phi^+ \phi) \;,
\end{eqnarray}
where $F=(\Sigma^0)^2+2\Sigma^+\Sigma^-$,  $b_4=\lambda_2+\lambda_1/2$ and $a_2=\alpha+\beta/2$. 
%\color{red}We emphasize that small $a_1$ term corresponds to a soft breaking of a global symmetry 
%$O(4)_H \times O(3)_\Sigma$ and the discrete symmetry $\Sigma \to -\Sigma$. 
We will work with the convention $a_1>0$ by absorbing the sign into the definition of $\Sigma$~\cite{FileviezPerez:2008bj}.
Note that for the potential to be bounded below, $\lambda_0$ and $b_4$ need to be larger than zero. This $b_4$ requirement   turns out to be very important which rules out a lot of parameter spaces for providing solution to the CDF $W$ mass excess problem. 

Minimization for the potential gives
\begin{eqnarray}
	 M^2_\Sigma &=&b_4 v_\Sigma^2- \frac{a_1 v^2}{4v_\Sigma}+\frac{a_2}{2}v^2\;,\nonumber\\
	m_\phi^2 &=&\lambda_0 v^2-\frac{a_1}{2}v_\Sigma +\frac{a_2}{2}v^2_\Sigma\;. 
\end{eqnarray}
Upon electroweak symmetry breaking,
mass eigenstates of the singly-charged states and  neutral states
are obtained by the corresponding mixing angles $\beta'_+$ and $\theta$
\begin{eqnarray}
	\left (\begin{array}{l}
		G^+\\
		H^+
	\end{array}
	\right ) = \left (\begin{array}{cc}  \cos\beta'_+\;\;\;&\; \;\; \sin\beta'_+\\  -\sin \beta'_+\;\;\;&\; \;\; \cos\beta'_+ \end{array} \right )
	\left (\begin{array}{l}
		\phi^+\\
		\Sigma^+
	\end{array}
	\right )\;,\;\;
	\left (\begin{array}{l}
		H_1\\
		H_2
	\end{array}
	\right ) = \left (\begin{array}{cc}  \cos\theta\;\;\;&\; \;\; \sin\theta\\  -\sin \theta\;\;\;&\; \;\; \cos\theta \end{array} \right )
	\left (\begin{array}{l}
		\phi^0\\
		\sigma
	\end{array}
	\right )\;,
\end{eqnarray}
where $G^+(\chi)$ are Nambu-Goldstone bosons which will be eaten by $W^+(Z)$  to provide the longitudinal components. Note that the physical fields are singly charged $H^+$,  neutral $H_1$ and $H_2$.

The mass squared matrix for charged fields  are
\begin{eqnarray}
	M_{\pm}^2 = a_1 \left (\begin{array}{cc}   v_\Sigma\;\;\;&\; \;\; \frac{v}{2}\\ \\
		\frac{v}{2}\;\;\;&\; \;\; \frac{ v^2}{4v_\Sigma} \end{array} \right )\;.
\end{eqnarray}
By diagonalizing the mass matrix with the mixing angle $\beta'_+$, 
the $H^\pm$ mass is
\begin{eqnarray}\label{chargedmass}
	m_{H^\pm}^2 =a_1 v_\Sigma \left(1+ \frac{v^2}{4v_\Sigma^2} \right) \;,\;\;\;
		\mbox{with}\;\; \tan 2\beta'_+=\frac{4v v_\Sigma}{4v_\Sigma^2-v^2}
\end{eqnarray}

The mass squared matrix for neutral fields  are
\begin{eqnarray}
	M_{0}^2 = \left (\begin{array}{cc}   2\lambda_0 v^2\;\;\;&\; \;\; -\frac{a_1}{2}v+a_2 v_\Sigma v\\ \\ -\frac{a_1}{2}v+a_2 v_\Sigma v\;\;\;&\; \;\;2b_4 v_\Sigma^2+\frac{a_1 v^2}{4v_\Sigma}
	 \end{array} \right )\;. 
\end{eqnarray}
By diagonalizing the mass matrix with the mixing angle $\theta$, we obtain the eigenvalues as 
\begin{eqnarray}\label{neutralmass}
	&&m^2_{H_1} =\lambda_0 v^2\left( 1+\frac{1}{\cos 2\theta} \right) + \left(b_4 v_\Sigma^2 +\frac{a_1 v^2}{8v_\Sigma} \right)\left( 1-\frac{1}{\cos 2\theta} \right) \;,\nonumber\\
	&&m^2_{H_2} =\lambda_0 v^2\left( 1-\frac{1}{\cos 2\theta} \right) + \left(b_4 v_\Sigma^2 +\frac{a_1 v^2}{8v_\Sigma} \right)\left( 1+\frac{1}{\cos 2\theta} \right) \;,\nonumber\\
&&	\mbox{with}\;\;\;
		\tan 2\theta= \frac{-a_1 v+2a_2 v_\Sigma v} {2\lambda_0 v^2-2b_4 v_\Sigma^2 -a_1 v^2/4v_\Sigma}
\end{eqnarray}
We identify the scalar $H_1$  with the SM-like Higgs $h$ and take the limit $\theta\approx 0$ to ensure its couplings consistent.  
We rename $H_2$ to be $H^0$. In total four scalars will be introduced,  the singly charged $H^{\pm}$, neutral CP-even $H^0$ and the SM-like Higgs $h$. Their effects on the oblique parameters can be parameterized  by the singly charged mass $m_{H^+}$, the mass difference $\Delta m=m_{H^+}-m_{H^0}$ and triplet vev $v_\Sigma$. The  limit $\theta \approx 0$ can be achieved by choosing $-a_1+2a_2 v_\Sigma \to 0$ for nonzero $v_\Sigma$. 
 In this case,
note that since $v>>v_\Sigma$, one finds $m_{H^+}\approx m_{H^0}$. This shows that the new scalars should be almost the degenerate states, which is consistent with the following  vacuum stability and perturbative unitarity of potential parameters $b_4$.
The mixing angle $\theta$ could be nonzero depending on the new Higgs mass. 
In our model from Eq.~\ref{neutralmass}, we can see that the leading order for $\theta$ is proportional to $v_\Sigma/v$, which is small. Therefore, the  $\theta$  effect will be small for the potential parameters and  physical mass of scalars.  %The effect can be avoided by choosing the physical mass as the fit parameters in the following. 

For one-loop level,  the oblique parameters  can be obtained by choosing the weak isospin $J=1$ and hypercharge $Y=0$ in Ref.~\cite{Lavoura:1993nq}
\begin{eqnarray}\label{loop1}
	S&=&	-\frac{2}{\pi}\sum_{I_3=-1}^{1}(I_3c_W^2)^2\xi\left(\frac{m^2_{I_3}}{m^2_{Z}},\frac{m^2_{I_3}}{m^2_{Z}}\right)=-\frac{4c_W^4}{\pi}\xi\left(\frac{m^2_{H^+}}{m^2_{Z}},\frac{m^2_{H^+}}{m^2_{Z}}\right) \approx -\frac{c_W^4}{15\pi}\frac{m_Z^2}{m^2_{H^+}}\;,\nonumber\\
	T&=&\frac{1}{16\pi c_W^2 s_W^2} \sum_{I_3=-1}^{1}(2-I_3^2+I_3) \eta\left(\frac{m^2_{I_3}}{m_Z^2},\frac{m^2_{I_3-1}}{m_Z^2}\right)
	= \frac{1}{12\pi c_W^2 s_W^2 m_Z^2}\frac{(m^2_{H^+}-m^2_{H^0})^2}{m^2_{H^+}}\;,\nonumber\\
	U&=&\frac{1}{6\pi } \sum_{I_3=-1}^{1}(2-3I_3^2)\ln \frac{m_{I_3}^2}{\mu^2}
	+\frac{1}{\pi}\sum_{I_3=-1}^1
	\left[ 2(I_3c_W^2)^2\xi\left(\frac{m^2_{I_3}}{m_Z^2},\frac{m^2_{I_3}}{m_Z^2}\right)
	-(2-I_3^2+I_3)\xi\left(\frac{m^2_{I_3}}{m_W^2},\frac{m^2_{I_3-1}}{m_W^2}\right)\right]\nonumber\\
	&=&\frac{1}{6\pi } \ln \frac{m_{H^0}^4}{m_{H^+}^4}+\frac{4}{\pi}
	\left[ c_W^4\xi\left(\frac{m^2_{H^+}}{m_Z^2},\frac{m^2_{H^+}}{m_Z^2}\right)
	-\xi\left(\frac{m^2_{H^+}}{m_W^2},\frac{m^2_{H^0}}{m_W^2}\right)\right]
	\approx \frac{2}{3\pi } \ln \frac{m_{H^0}}{m_{H^+}}-\frac{c_W^2 s_W^2}{15\pi}\frac{m_Z^2}{m^2_{H^+}}\;.
\end{eqnarray}
Here $m_{\pm 1}=m_{H^\pm}$ and $m_{0}=m_{H^0}$. 
Our results are more general compared to the ones in Ref.~\cite{Khan:2016sxm} because  we contain   the high order term $O(1/m_{H^+}^2)$. 

The tree level contribution has been analyzed in Ref.~\cite{Cheng:2022aau,FileviezPerez:2022lxp} previously.  For a complete analysis the overall triplet contributions to the oblique  parameters should contain  the tree-level one in Eq.~(\ref{tree1}) and one-loop full expression in Eq.~(\ref{loop1}) simultaneously.  The investigations carried out in~\cite{Cheng:2022aau,FileviezPerez:2022lxp} have partially carried out the analysis. Here we fill in the gap to make such a complete analysis.
\\

\section{ Electroweak global fits}

To assess the impact of the new measurement $m_W$ on the potential NP implications, we   perform the global fit from the Bayesian analysis with the   HEPfit code~\cite{DeBlas:2019ehy} and all inputs data used are shown in Table-\ref{data}. The same input parameters are  also shown  in Ref.~\cite{Lu:2022bgw}.

\begin{table}[!htb]
	\caption{The input parameters and the best points in the global fit.  The Fermi constant $G_F=1.1663787\times 10^{-5} [\mbox{GeV}^{-2}]$.  $m_h=125.25$ GeV, $m_t=172.76$ GeV, $m_b=4.18$ and $m_c=1.28$ GeV are used as the SM reference point.}
	\label{data}
	\begin{tabular}{|c|c|c|c|c|c|c|c|c|c|}
		\hline \hline
		\multirow{2}{*}{parameter} &\multirow{2}{*}{Input Value}& STU & ST & Y=1&  Y=0 \tabularnewline\cline{3-6}
		& &$\chi_{min}^2(\mbox{dof})=13.76(13) $&$\chi_{min}^2(\mbox{dof})=17.20 (14)$ & $\chi_{min}^2(\mbox{dof})=16.04 (13)$ &$\chi_{min}^2(\mbox{dof})=21.22(13) $\tabularnewline
		\hline
		$m_W[{\rm GeV}]$&$80.4335(94)$& 80.434(9) &80.428(9)&80.431(9)&80.417(8)\tabularnewline\hline
		$\Delta\alpha^{(5)}_{\rm had}$&$0.02761(11)$&0.02761(8) & 0.027614(78)&0.02761(8)&0.027619(77)\tabularnewline\hline
%	$m_h[{\rm GeV}]$&$125.25(17)$&125.25(0)&125.25(0)&&\tabularnewline\hline
	%	$m_t[{\rm GeV}]$&$172.76(58)$&   172.76(0) &172.76(0) &&\tabularnewline\hline
		$\alpha_s(m_Z)$&$0.1179(9)$& 0.11788(62)&0.11775(62)&0.11777(62)&0.11776(61)\tabularnewline\hline
		$\Gamma_W[{\rm GeV}]$&$2.085(42)$&2.0941(8)&2.0936(7) &2.0938(7)&2.0927(7)\tabularnewline\hline
		$\Gamma_Z[{\rm GeV}]$&$2.4952(23)$&  2.4958(22)& 2.4997 (9) &2.4991(6)&2.4980(5)\tabularnewline\hline
		$m_Z[{\rm GeV}]$&$91.1875(21)$&91.187(1) & 91.187(1) &91.187(1)&91.188(1)\tabularnewline\hline
		$A^{0,b}_{FB}$&$0.0992(16)$& 0.10311(83) &0.10346(79) &0.10302(32)&0.10468(28)\tabularnewline\hline
		$A^{0,c}_{FB}$&$0.0707(35)$& 0.073667(639)&0.073937(610)&0.073594(251)&0.074877(215)\tabularnewline\hline
		$A^{0,\ell}_{FB}$&$0.0171(10)$&0.016216(256)&0.016324(244)&0.016186(100)&0.0167(1)\tabularnewline\hline
		$A_b$&$0.923(20)$&0.93474(10)&0.93478(9) &0.93473(4)&0.93492(3)\tabularnewline\hline
		$A_c$&$0.670(27)$&  0.66781(51)&0.66803(49)&0.66776(20)&0.66878(17)\tabularnewline\hline
		$A_\ell({\rm SLD})$&$0.1513(21)$&0.147(1) &0.14749(111) &0.14687(46)&0.14921(39)\tabularnewline\hline
		$A_\ell({\rm LEP})$&$0.1465(33)$&0.147(1) &0.14749(111)&0.14687(46)&0.14921(39)\tabularnewline\hline
		$R^0_b$&$0.21629(66)$&0.21597(1) &0.21597(1)&0.21597(1)&0.21596(7)\tabularnewline\hline
		$R^0_c$&$0.1721(30)$& 0.17226(1)&0.17226(1)&0.17225(1)&0.17227(1)\tabularnewline\hline
		$R^0_\ell$&$20.767(25)$& 20.755(5)&20.755(5)&20.754(4)&20.759(4)\tabularnewline\hline
		$\sigma^0_h[{\rm nb}]$&$41.540(37)$&41.498(3) &41.499(3) &41.499(3)&41.497(3)\tabularnewline\hline
		$\sin^2\theta^\ell_{\rm eff}(Q_{\rm FB})$&$0.2324(12)$&0.23151(15) &0.23145(14) &0.23153(6)&0.23123(5)\tabularnewline\hline
		$\sin^2\theta^\ell_{\rm eff}({\rm Teva})$&$0.23148(33)$&0.23151(15)&0.23145(14)&0.23153(6)&0.23123(5)\tabularnewline\hline
	%	$\overline{m_c}[{\rm GeV}]$&$1.27(2)$& 1.28(0)&1.28(0)&&\tabularnewline\hline
	%	$\overline{m_b}[{\rm GeV}]$&$4.18^{(3)}_{(2)}$& 4.18(0)&4.18(0)&&\tabularnewline\hline
	\end{tabular}
\end{table}

The theory uncertainties we use are~\cite{deBlas:2021wap,deBlas:2022hdk}
\begin{eqnarray}
	&&	\delta_{th} m_W=4 \mbox{MeV}\;,\; \;\;\delta_{th} s^2_W=5\times 10^{-5}\;,\;\;\;
	\delta_{th} \Gamma_Z=0.4 \mbox{MeV}\;,\;\;\; \delta_{th} \sigma^0_{had} =6 \mbox{pb}\;,\;\;\;\nonumber\\
	&&\delta_{th} R^0_l=0.006\;,\;\;\;\delta_{th} R^0_l=0.00005\;,\;\;\;\delta_{th} R^0_b=0.0001\;
.
\end{eqnarray}

In view of the significant discrepancy between  CDF $W$ mass measurement and the SM prediction, we  use two different ways,  the oblique parameters and triplet model parameters,  to conduct the global fits  and further compare these two corresponding results.

\begin{table}[!htb]
	\caption{The fit parameter values for $S$, $T$ and $U$ and the correlation matrix.}
	\label{table2}
	\begin{tabular}{|c|c|ccc|c|c|cc|}
		\hline \hline
		13 dof & Result   &  \multicolumn{3}{|c|}{Correlation} 
		&14 dof & Result   & \multicolumn{2}{|c|}{Correlation} \tabularnewline
		& $\chi^2_{min}=13.76$   & $S$&$T$& $U$
		&& $\chi^2_{min}=17.20$   & $S$&$T$ \tabularnewline 
		\hline
		$S$ &$0.03 \pm 0.03$ & 1&0.9&-0.63
		&	$S$ &$0.14 \pm 0.03$ & 1&0.93\tabularnewline
		\hline
		$T$   &$0.06 \pm 0.02$ & &1&-0.87 
		&	$T$   &$0.24 \pm 0.02$ & &1 \tabularnewline
		\hline
		$U$  &$0.16 \pm 0.03$ & &&1
		& $U=0$  & & & \tabularnewline
		\hline
	\end{tabular}
\end{table}

\begin{figure}[!t]
	\centering
	\subfigure[\label{ST}. S-T contour]
	{\includegraphics[width=.4\textwidth]{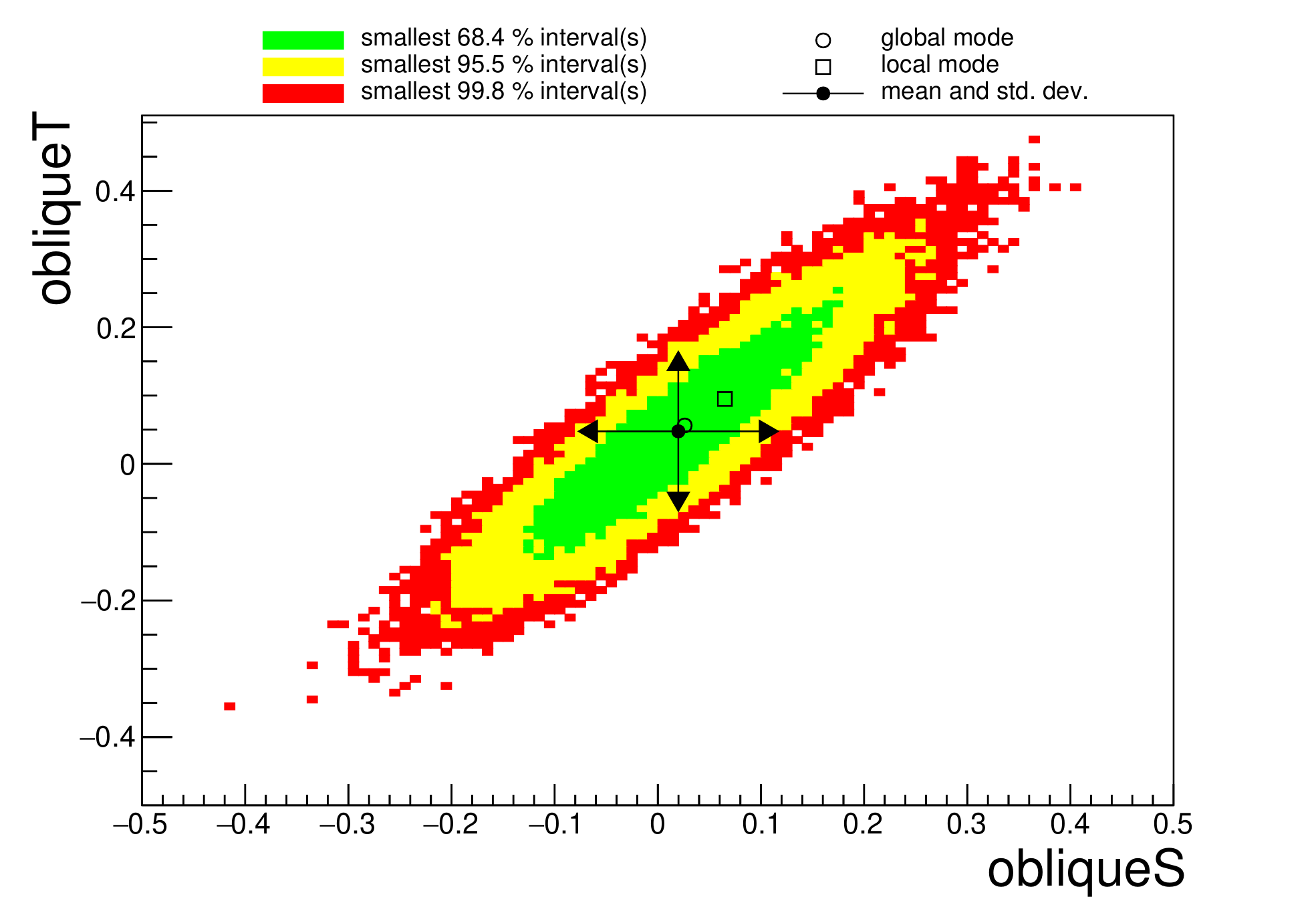}}
	\subfigure[\label{SU}. S-U contour]
	{\includegraphics[width=.4\textwidth]{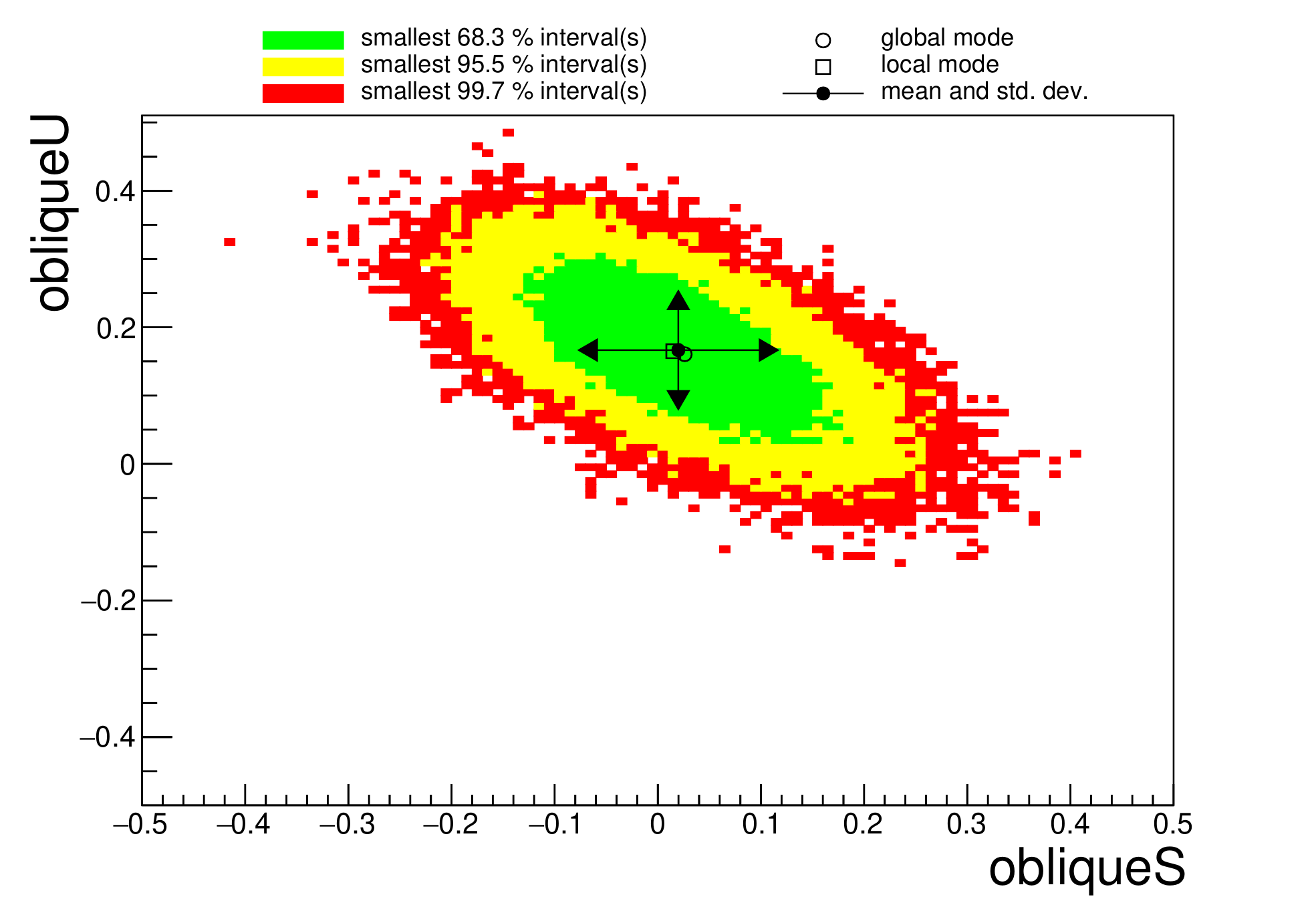}}
	\subfigure[\label{TU}. T-U contour]
	{\includegraphics[width=.4\textwidth]{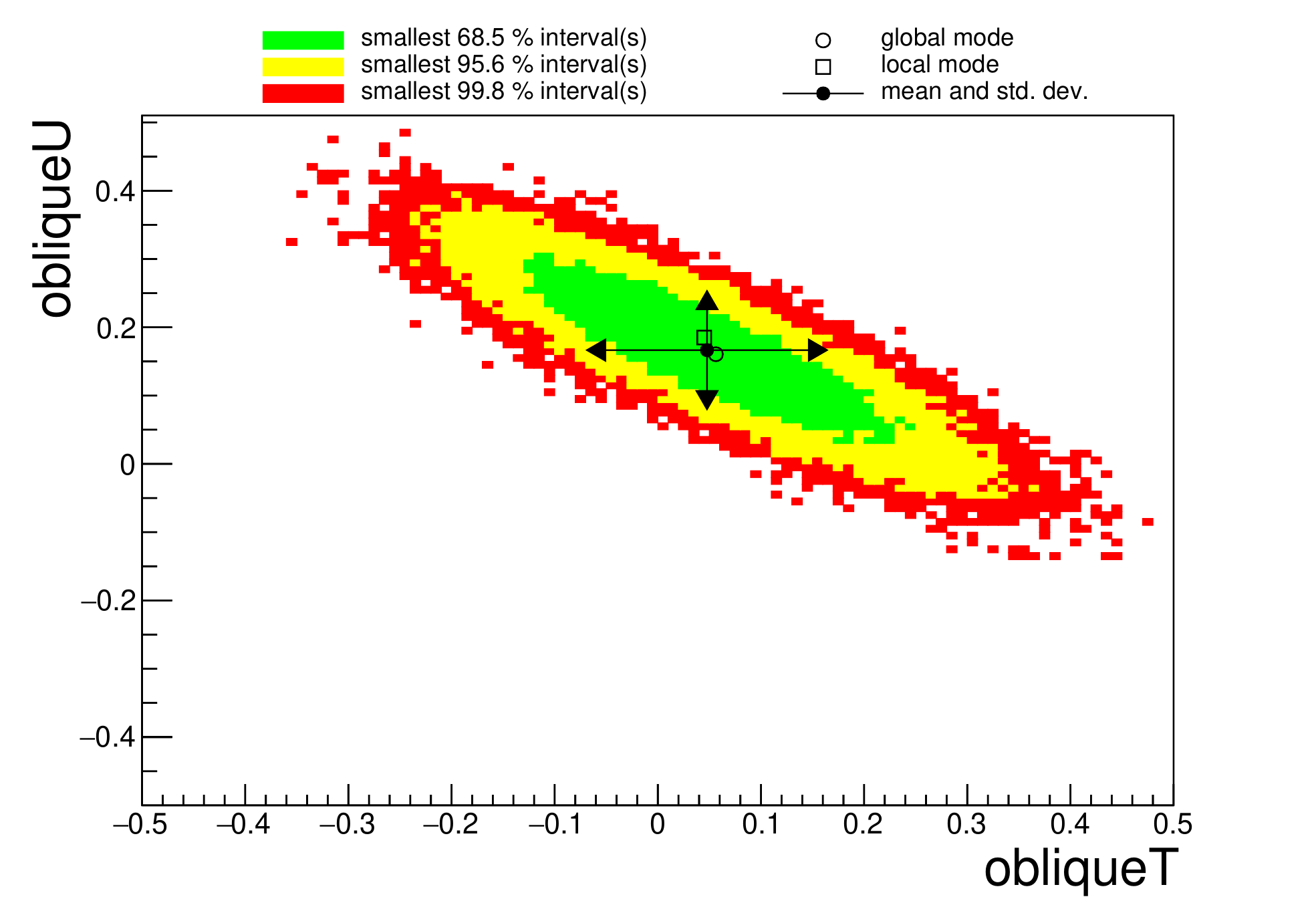}}
	\subfigure[\label{STU0}. S-T contour with $U=0$]
	{\includegraphics[width=.4\textwidth]{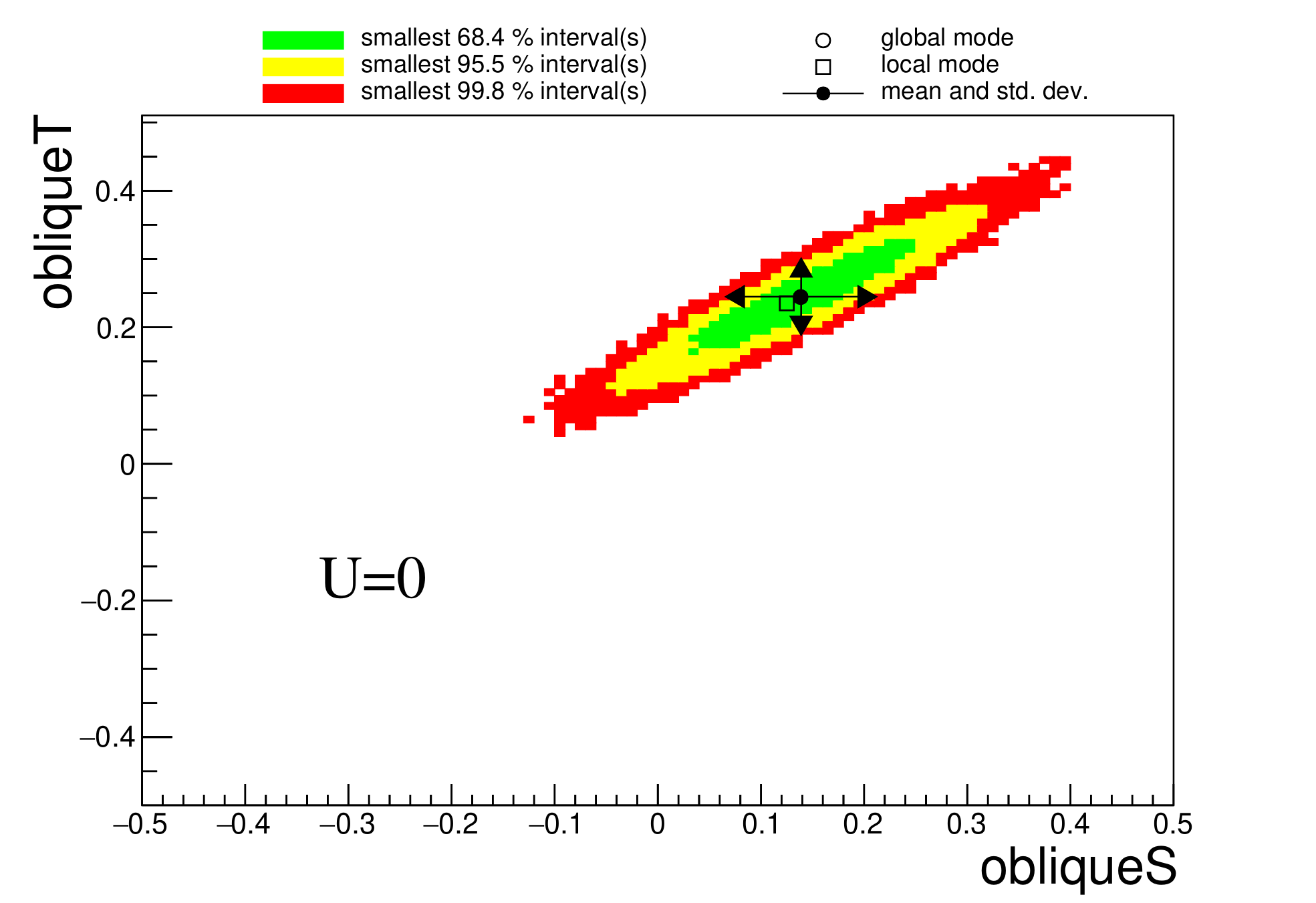}}
	\caption{ Contour plot for oblique parameters $S$, $T$ and $U$. The black dot denotes the best fit value, $1\sigma$ in green, $2\sigma$ in yellow and $3\sigma$ in red, respectively.}
	\label{STU}
\end{figure}

Let us first consider the case of oblique corrections.  In terms of the $S$, $T$ and $U$ parameters, we  obtain $S=0.03\pm 0.03$, $T=0.06\pm 0.02$ and $U=0.16\pm0.03$ with $\chi^2_{min}/d.o.f=13.76/13$  in the left panel of Table-\ref{table2}. 
%Here the number in the bracket denotes the degree of freedom ($d.o.f$). 
Our results are consistent within $1\sigma$ with Ref.~\cite{Carpenter:2022oyg,Lu:2022bgw,deBlas:2022hdk,Asadi:2022xiy}. The corresponding contour plots are shown in Fig.~\ref{STU}, the best fit in black dot, $1\sigma$ in green, $2\sigma$ in yellow, and $3\sigma$ in red.

This scenario, $U<<S,T$, is expected in extensions with heavy new physics where the SM gauge symmetries are realized linearly in the light fields, in which case $U$  generated by mass dimension-8 is suppressed with respect to $S$ and $T$ from dimension-6 interactions. Thus, one can assume $U=0$. Then we  obtain $S=0.14\pm 0.03$ and $T=0.24\pm 0.02$  with $\chi^2_{min}/d.o.f=17.20/14$  in the right panel of Table-\ref{table2}. Our results are consistent within $1\sigma$ with Ref.~\cite{Strumia:2022qkt,Paul:2022dds,Balkin:2022glu,Carpenter:2022oyg,Lu:2022bgw,deBlas:2022hdk,Asadi:2022xiy,Gu:2022htv,Liu:2022vgo}. Similarly, the corresponding contour plot is shown in Fig.~\ref{STU0}, the best fit in black dot, $1\sigma$ in green,  $2\sigma$ in yellow, and $3\sigma$ in red.

If further adopting $S=U=0$, we can obtain the fit results $T=0.15\pm 0.02$ with $\chi^2_{min}/d.o.f=20.6/15$. We find that  much smaller $T$ parameter compared to the $S \neq 0$ case can  reasonably solve the CDF excess. This can be  understood that $S$ and $T$ parameters contribute to  $W$ mass with the opposite sign.

Compared with the global fit results without the CDF new data~\cite{pdg},  $S= -0.02\pm 0.10$, $T=0.03\pm0.12$ and $U=0.01\pm 0.11$ (or $S=-0.01\pm 0.07$, $T=0.04\pm 0.06$ with $U=0$), one can see that the CDF new data has remarkable impact on the results. The inclusion of CDF new data in the analysis indicates new physics beyond SM in a significantly way. When  applied to a given model, the resulting best fit values for $S$, $T$ and $U$ parameters may be not the same as taking them to be independent ones since they are correlated in specific models. To this end, we carried out global fits by using the triplet model parameters directly presented earlier. 

\begin{figure}[!t]
	\centering
	\subfigure[\label{mHvdelta} $m_{H^{++}}-v_\Delta$ contour]
	{\includegraphics[width=.4\textwidth]{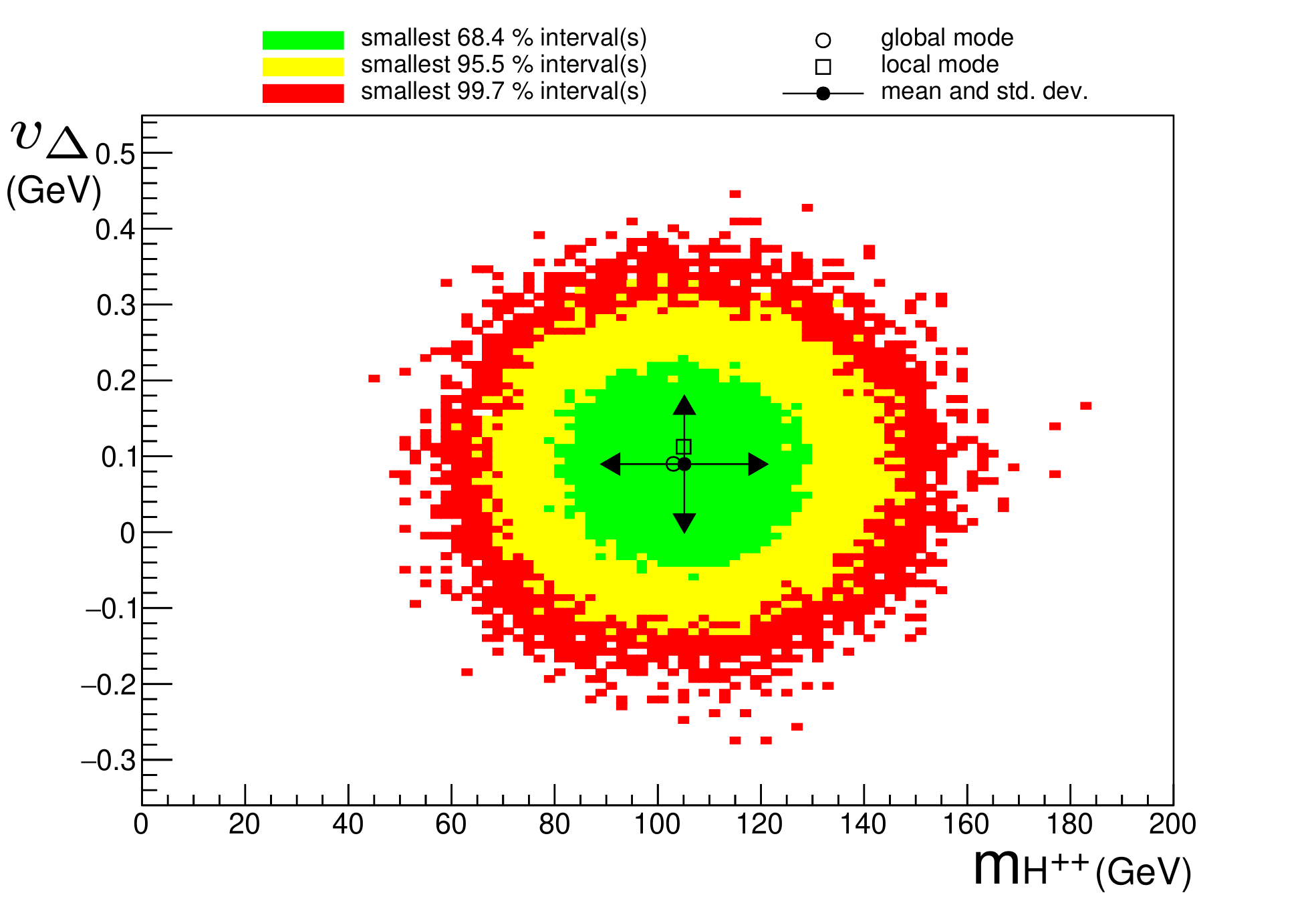}}
	\subfigure[\label{mHlambda4} $m_{H^{++}}-\lambda_4$ contour]
	{\includegraphics[width=.4\textwidth]{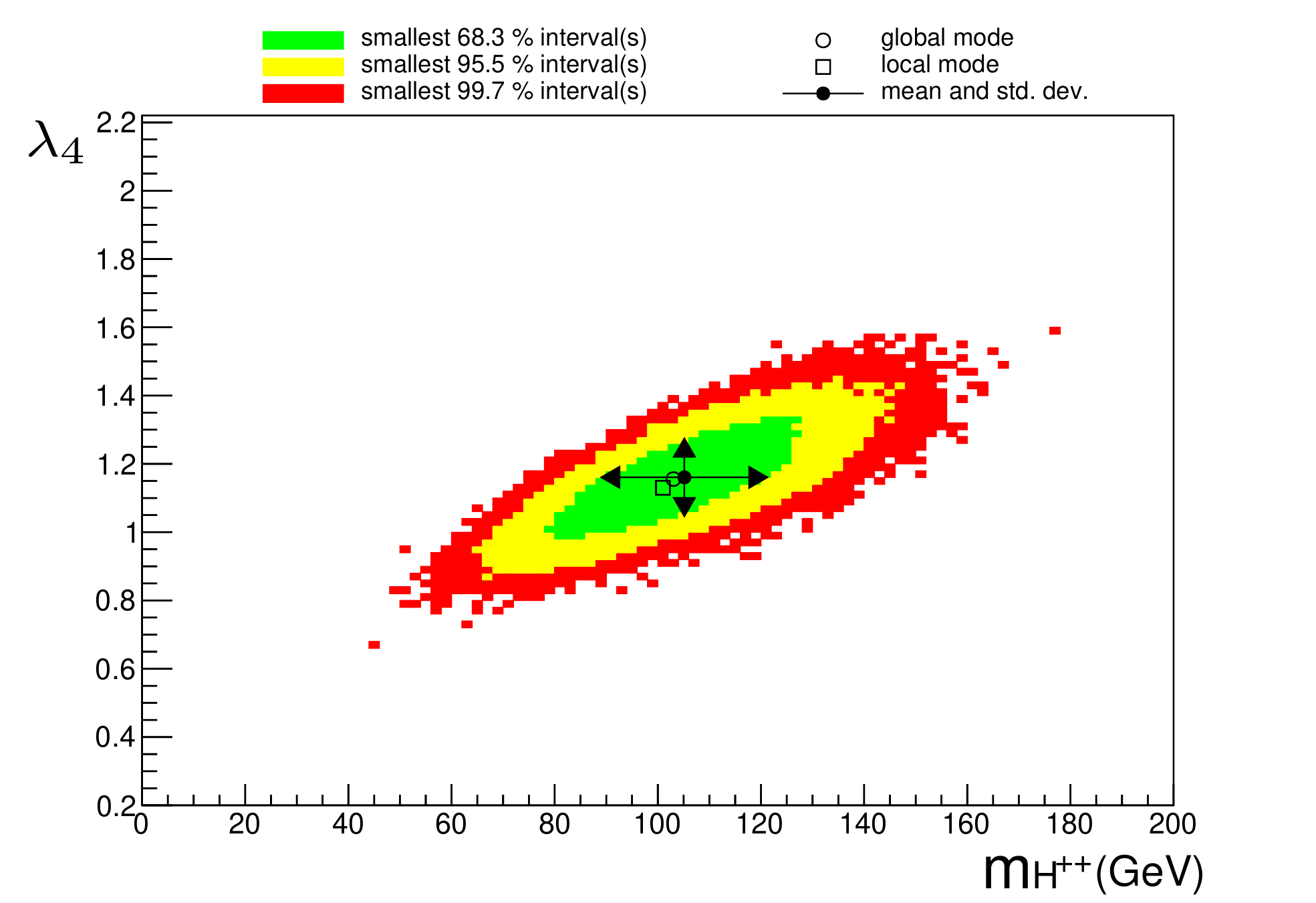}}
	\subfigure[\label{lambda4vdelta} $\lambda_4-v_\Delta$ contour]
	{\includegraphics[width=.4\textwidth]{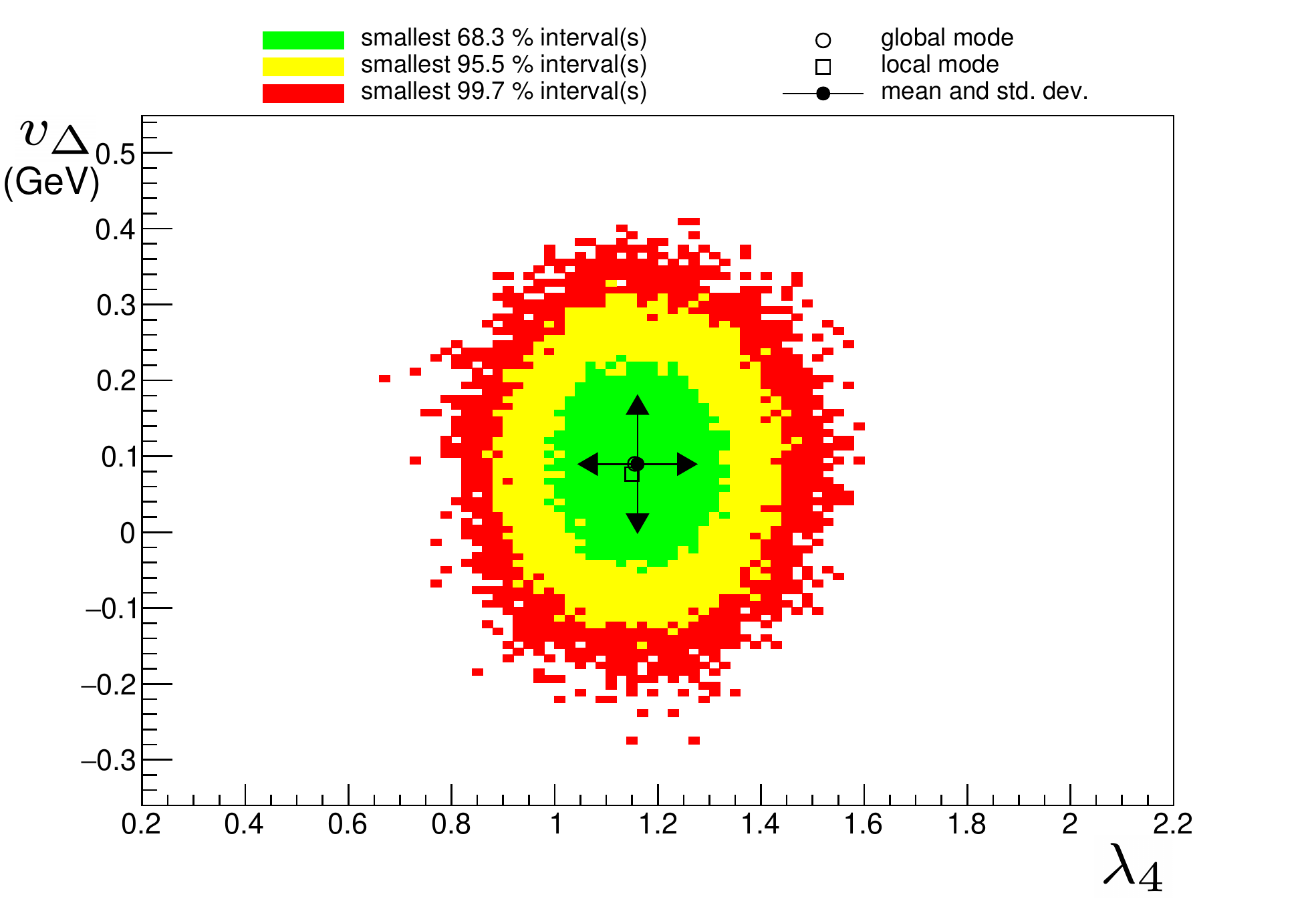}}
	\caption{ Contour plot for the fit parameters $m_{H^{++}}$, $\lambda_{4}$ and $v_{\Delta}$ for $Y=1$ scalar. The black dot denotes the best fit value, $1\sigma$ in green, $2\sigma$ in yellow and $3\sigma$ in red, respectively.}
	\label{Y1}
\end{figure}

For the $Y=1$ triplet  case,  we express the parameters $S$, $T$ and $U$ in terms of $m_{H^{++}}$, $\lambda_4$ and $v_\Delta$ by combining the tree-level in Eq.~(\ref{tree}) and one-loop full expressions in  Eq.~(\ref{loop}). 
Using the three new triplet parameters to perform the global fit, we obtain $m_{H^{++}}$, $\lambda_4$ and $v_\Delta$ with reasonable $\chi^2_{min}/d.o.f =16.04/13$ as shown in Table-\ref{table3}.  The corresponding contour plot is shown in Fig.~\ref{Y1}, the best point in black dot, $1\sigma$ in green, $2\sigma$ in yellow, and $3\sigma$ in red. 
This further results in $\Delta m=m_{H^+}-m_{H^{++}}=64.78\pm 2.39$ GeV. 
We find that our fit values are consistent  with the results in Ref.~\cite{Heeck:2022fvl}. And $v_\Delta$ is approximately zero, which shows the tree level contribution is pretty small. Furthermore, using our fit results, we obtain the central values of the oblique parameters,  $S=0.15$, $T=0.22$, $U=0.04$, which are consistent with values obtained in the right panel of Table-\ref{table2} treating $S$, $T$ and $U=0$ independently.

\begin{table}[!htb]
	\caption{The fit values  $m_H$, $\lambda_4$ and $v_\Delta$ and the correlation matrix in $Y=1$ scalar.}
	\label{table3}%
	\begin{tabular}{|c|c|ccc|}
		\hline \hline
		13 dof & Result   &  \multicolumn{3}{|c|}{Correlation}  \tabularnewline
		& $\chi^2_{min}=16.04$   & $m_{H^{++}}({\rm GeV})$&$\lambda_4$&$v_\Delta({\rm GeV})$  \tabularnewline
		\hline
		$m_{H^{++}}({\rm GeV})$ &$103.02 \pm 9.84$ &1&0.78&0.1\tabularnewline
		\hline
		$\lambda_4$   &$1.16 \pm 0.07$  &&1&0.056\tabularnewline
		\hline
		$v_\Delta({\rm GeV})$  &$0.09 \pm 0.09$ &&&1\tabularnewline
		\hline
	\end{tabular}
\end{table}

\begin{figure}[!t]
	\centering
	\subfigure[\label{mHvsigma} $m_{H^+}-v_\Sigma$ contour]
	{\includegraphics[width=.4\textwidth]{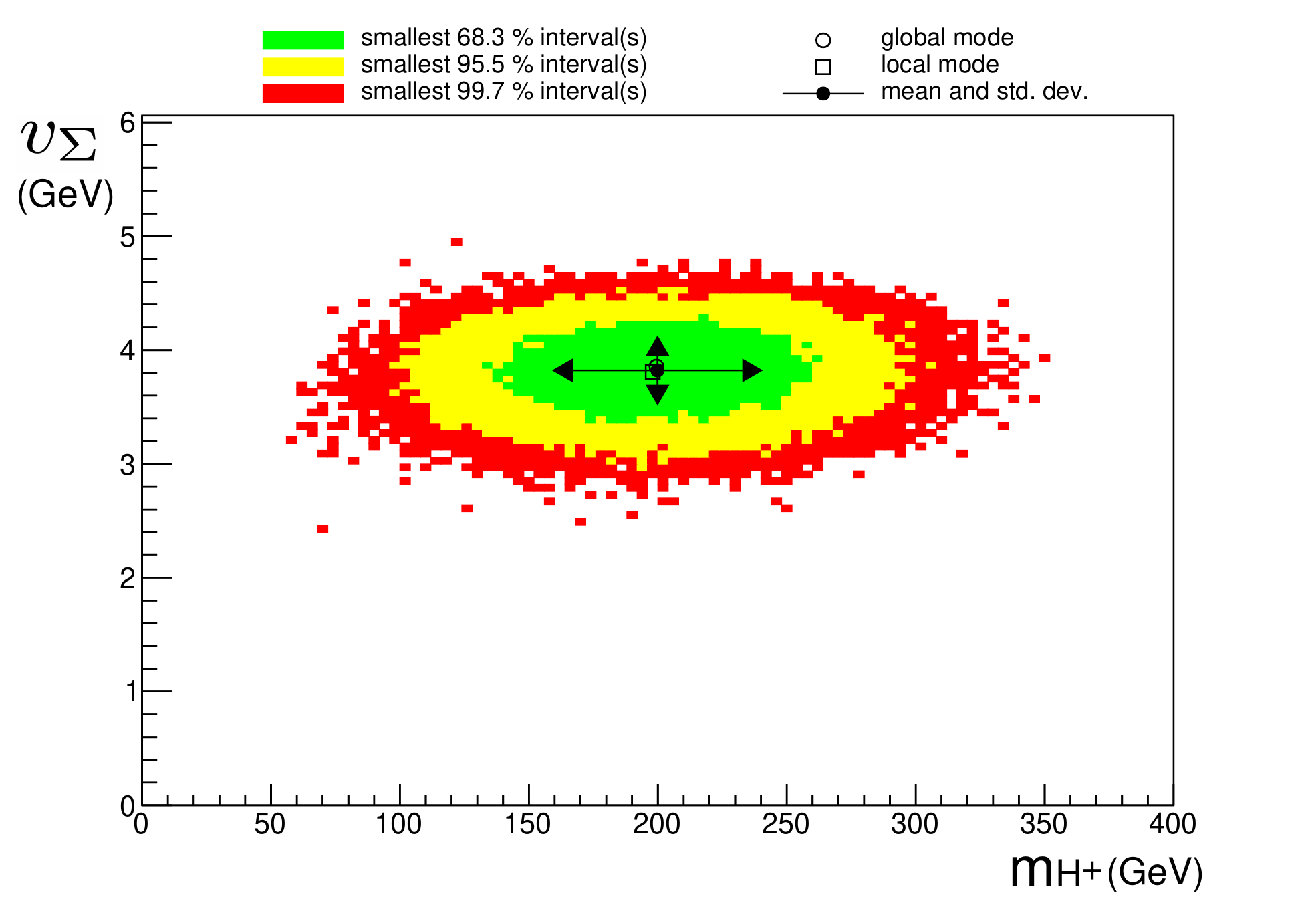}}
	\subfigure[\label{mHdeltam} $m_{H^+}-\Delta m$ contour]
	{\includegraphics[width=.4\textwidth]{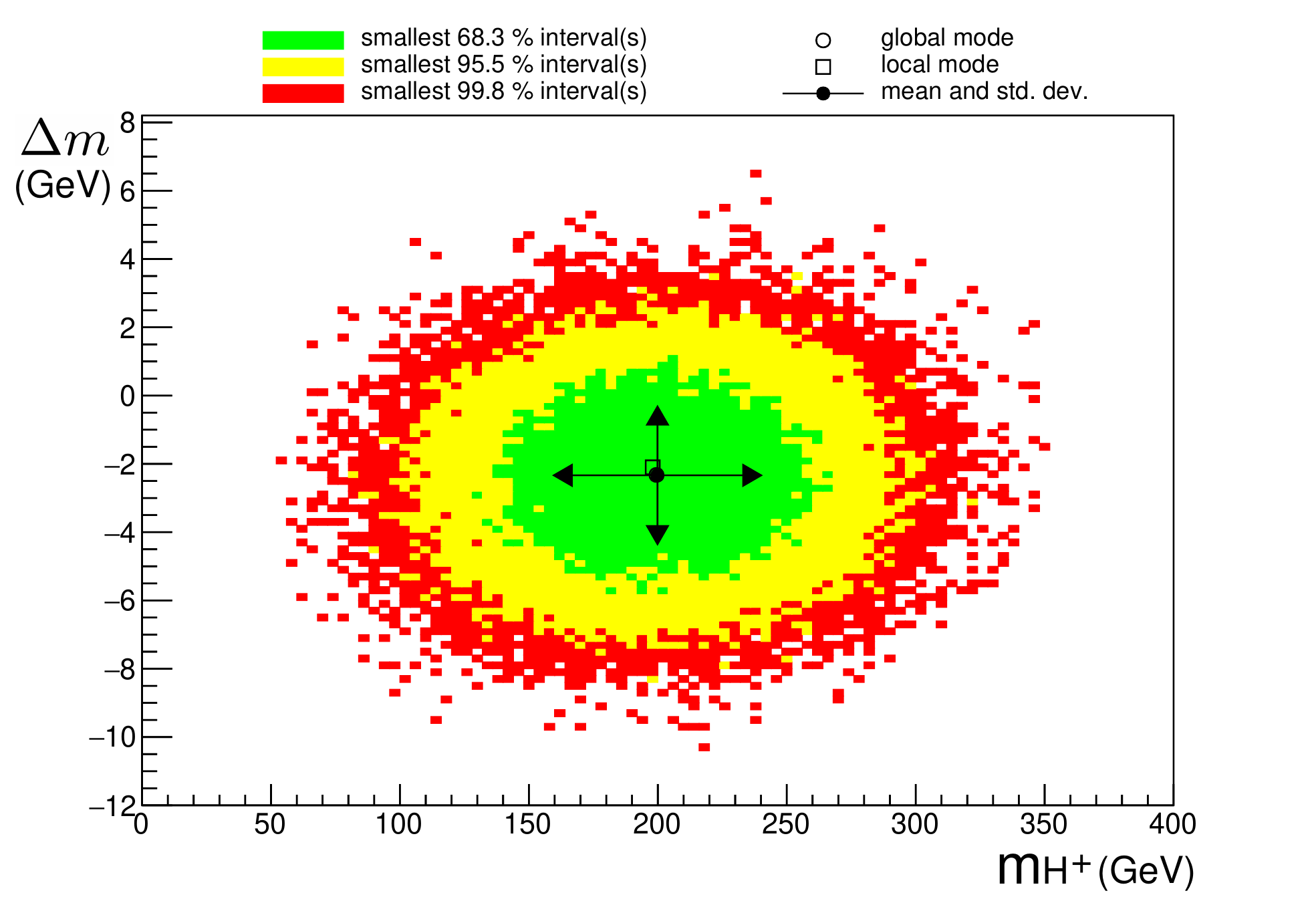}}
	\subfigure[\label{deltamvsigma} $\Delta m-v_\Sigma$ contour]
	{\includegraphics[width=.4\textwidth]{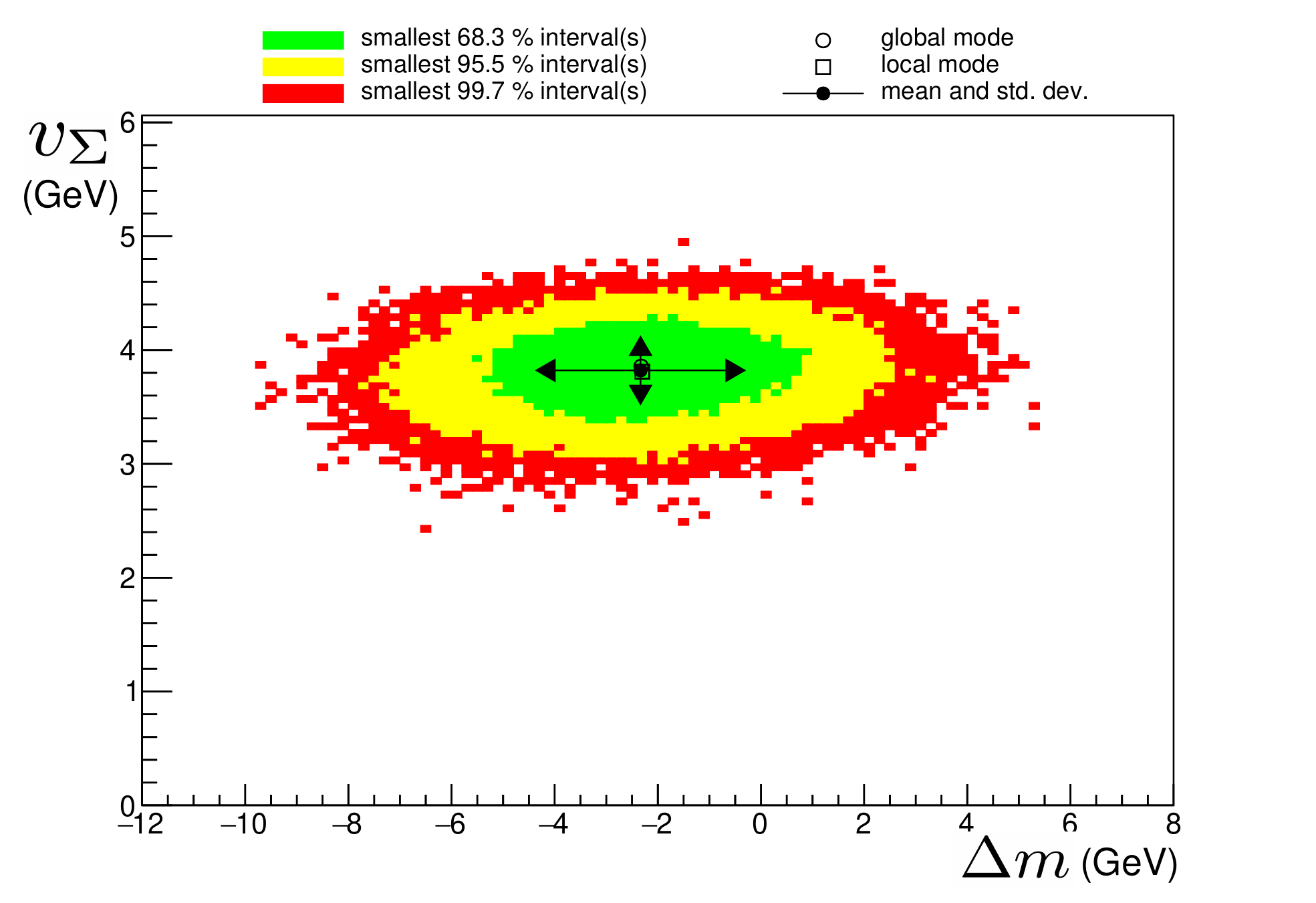}}
	\caption{ Contour plot for fit parameters $m_{H^+}$, $\Delta m$ and $v_{\Sigma}$ for $Y=0$ scalar. The black dot denotes the best fit value, $1\sigma$ in green, $2\sigma$ in yellow and $3\sigma$ in red, respectively.}
	\label{Y0}
\end{figure}

For the $Y=0$ triplet case,  we similarly express the  parameters $S$, $T$ and $U$ in terms of $m_{H^+}$, $\Delta m$ and $v_\Sigma$ by combining tree-level in Eq.~(\ref{tree1}) and one-loop full expression in  Eq.~(\ref{loop1}). Using the three  parameters to perform the global fit, we obtain the fit results as shown in Table-\ref{table4}. The corresponding contour plot is shown in Fig.~\ref{Y0}, the best fit in black dot, $1\sigma$ in green, $2\sigma$ in yellow, and $3\sigma$ in red.  
We find that $v_\Sigma $ is around several GeVs, which is totally different from $Y=1$ case.   This is because the tree level contribution in $Y=0$ case can give the right direction to explain CDF $W$ measurement.  And  if $v_\Sigma=0$, $\phi$ and $\Sigma$ do not mix and  tree level masses are degenerate. Radiative corrections break the degeneracy between the charged and neutral components of the triplet only around $\Delta m \sim O({\rm MeV})$ as shown in Ref.~\cite{FileviezPerez:2008bj}. These both indicate that non-zero vev is needed. 
%Similarly, if setting $v_\Sigma=0$ to perform the global fit, we can obtain $m_{H}=64.76{\rm GeV}$, $\Delta m=-26.28{\rm GeV}$ with $\chi^2(d.o.f)=17.76(14)$.
 %In this case, there exists only the loop contribution. Although this case of $v_\Sigma=0$ has a reasonable  $\chi^2/d.o.f$, the fit results  are unreasonable. This is because that for $v_\Sigma=0$, $\phi$ and $\Sigma$ do not mix and  tree level masses are degenerate. Radiative corrections break the degeneracy between the charged and neutral components of the triplet around $\Delta m \sim O({\rm MeV})$ as shown in Ref.~\cite{FileviezPerez:2008bj}. 
Using our fit results, we obtain the central values of the oblique parameters,  $S=-0.003$, $T=0.13$, $U=0.002$. Note that  $U$ is not much smaller than $S$, which clearly is contradictory to the expectation that $U$ is negligible compared with $S$ and $T$ in the NP models. 
This  shows potentially that the assumption of $U<<S,T$ in $Y=0$ case may be improper.

\begin{table}[!htb]
	\caption{The fit parameter values for $m_{H^+}$, $\Delta m$ and $v_\Sigma$ and the correlation matrix in $Y=0$ scalar.}
	\label{table4}
	\begin{tabular}{|c|c|ccc|}
		\hline \hline
		13 dof & Result   & \multicolumn{3}{|c|}{Correlation} \tabularnewline
		& $\chi^2_{min}=21.22$   & $m_{H^+}({\rm GeV})$&$\Delta m({\rm GeV})$&$v_\Sigma({\rm GeV})$  \tabularnewline
		\hline
		$m_{H^+}({\rm GeV})$ &$199.45 \pm 39.73$ &1&0.031&0.056\tabularnewline
		\hline
		$\Delta m({\rm GeV})$   &$-2.32\pm1.99$  &&1&0.1\tabularnewline
		\hline
		$v_\Sigma({\rm GeV})$  &$3.86\pm 0.27$ && &1\tabularnewline
		\hline
	\end{tabular}
\end{table}

\section{ phenomenology}

\subsection{Discussions for the $Y=1$ triplet model}

Using the fit values in Table-\ref{table3} for $Y=1$ case, we further obtain $m_{H^{+}}=(167.82 \pm 8.73) \mbox{GeV}$, $m_{H}=m_{A}=(213.81\pm 9.15)\; \mbox{GeV}$.  
We find that 
 the fit central value  $m_{H^{++}}$ satisfies the current experimental ranges $84<m_{H^{++}}<200$ GeV as discussed earlier.
Furthermore, our results are in favor of the hierarchical mass spectrum $m_H>m_{H^{+}}>m_{H^{++}}$, which is consistent with Ref.~\cite{Kanemura:2012rs}.\\

\noindent
{\bf Vacuum stability and perturbative unitarity}

We can easily express the potential parameters $\mu$ and the $\lambda_i$ in terms of the physical scalar masses and the mixing angle $\alpha$  as well as the vev's $v,\;v_\Delta$ \cite{Arhrib:2011uy}. In the very small mixing angle $\alpha$, we  further calculate these values as
\begin{eqnarray}
	\lambda_1 &=& -\frac{2}{v^2+4v_\Delta^2}m_A^2
	+\frac{4}{v^2+2v_\Delta^2}m_{H^+}^2+\frac{\sin 2\alpha}{2v v_\Delta}(m_h^2-m_H^2)\approx 0.35\pm 0.07\;,\nonumber\\
	\lambda_2 &=& \frac{1}{v_\Delta^2} \left[ \frac{\sin^2 \alpha \;m_h^2+\cos^2 \alpha \;m_H^2}{2}+\frac{m_A^2}{2}\frac{v^2}{v^2+4v_\Delta^2}-\frac{2v^2}{v^2+2v_\Delta^2}
   m_{H^+}^2 +m^2_{H^{++}}	\right] \approx 1.86\pm 0.19\;,\nonumber\\
   	\lambda_3 &=& \frac{1}{v_\Delta^2} \left[ -\frac{v^2}{v^2+4v_\Delta^2}m_A^2+\frac{2v^2}{v^2+2v_\Delta^2}
   m_{H^+}^2 -m^2_{H^{++}}	\right] \approx -1.86\pm 0.19\;,\nonumber\\
   	\lambda_4 &=& \frac{4}{v^2+4v_\Delta^2} m_A^2 -\frac{4}{v^2+2v_\Delta^2}
   m_{H^+}^2	 \approx 1.16\pm 0.07\;,\nonumber\\
   	\lambda &=& \frac{2}{v^2} (\cos^2 \alpha\; m_h^2+\sin^2 \alpha m_H^2) 
  	 \approx 0.52\pm 0.0000004\;,\nonumber\\
   	\mu &=& \frac{\sqrt{2}v_\Delta}{v^2+4v_\Delta^2} m_A^2 	 \approx (0.10\pm 0.10) \;\mbox{GeV}\;.
\end{eqnarray}
We find that the above potential parameters fully satisfy the vacuum stability, which means that the scalar potential remains bounded from below in any directions of field space  \cite{Primulando:2019evb},
\begin{eqnarray}
&&	\lambda \ge 0, \;\;\lambda_2+\frac{\lambda_3}{2}\ge 0,\;\;\lambda_1+\sqrt{\lambda(\lambda_2+\lambda_3)}\ge 0,\;\;
	\lambda_1+\lambda_4+\sqrt{\lambda(\lambda_2+\lambda_3)}\ge 0\nonumber\\
&&	|\lambda_4|\sqrt{\lambda_2+\lambda_3}-\lambda_3 \sqrt{\lambda}\ge 0 \;\; or \;\;
2\lambda_1+\lambda_4 \sqrt{(2\lambda-\lambda_4^2/\lambda_3)(2\lambda_2+\lambda_3)} \ge 0.
\end{eqnarray}
Similarly,  the above potential parameters can also contribute to  the tree-level two-body scattering with required amplitudes  to be unitarity in all orders of perturbation calculation \cite{Arhrib:2011uy}. The following perturbative unitarity conditions are also preserved
\begin{eqnarray}
	&&	|(\lambda+4\lambda_2+8\lambda_3)\pm \sqrt{(\lambda-4\lambda_2-8\lambda_3)^2+16\lambda_4^2}| \le 64\pi \;,
 \nonumber\\
 	&&	|(3\lambda+16\lambda_2+12\lambda_3)\pm \sqrt{(3\lambda-16\lambda_2-12\lambda_3)^2+24(2\lambda_1+\lambda_4)^2}| \le 64\pi\;,
 \nonumber\\
 &&	 \lambda \le 32\pi, \;\;|2\lambda_1+3\lambda_4|\le 32\pi, \;\;|2\lambda_1-\lambda_4|\le 32\pi,\;\;|\lambda_1|\le 16\pi,\nonumber\\
  &&	| \lambda_1+\lambda_4| \le 16\pi, \;\;|2\lambda_2-\lambda_3|\le 16\pi, \;\;|\lambda_2|\le 8\pi,\;\;|\lambda_2+\lambda_3|\le 8\pi\;.
\end{eqnarray}

The above analysis is at the electroweak scale. To ensure the stability of potential at any high scale one should consider the renormalization group evolution. Henceforth, we further study the  renormalization group equation (RGE) for quartic couplings $\lambda_i$'s~\cite{Chun:2012jw,Bonilla:2015eha} as the following 
\begin{eqnarray}
		(4\pi)^2 \frac{d \lambda}{d t}&=&\frac{27}{50} g_1^4 +\frac{9}{5}g_1^2 g_2^2+\frac{9}{2}g_2^4 -\left(\frac{9}{5}g_1^2+9 g_2^2\right) \lambda +6\lambda^2+12\lambda_1^2+12\lambda_1\lambda_4+5\lambda_4^2+12\lambda y_t^2-24y_t^4\;,	\nonumber\\
		(4\pi)^2 \frac{d \lambda_1}{d t}&=&\frac{27}{25} g_1^4 -\frac{18}{5}g_1^2 g_2^2+6g_2^4 -\left(\frac{9}{2}g_1^2+\frac{33}{2} g_2^2\right) \lambda_1 +3\lambda \lambda_1+\lambda\lambda_4+4\lambda_1^2  \nonumber \\ 
	&+&16\lambda_1\lambda_2+12\lambda_1\lambda_3+\lambda_4^2+6\lambda_2\lambda_4+2\lambda_3\lambda_4+6\lambda_1 y_t^2\;,	\nonumber\\
			(4\pi)^2 \frac{d \lambda_2}{d t}&=&\frac{54}{25} g_1^4 -\frac{36}{5}g_1^2 g_2^2+15 g_2^4 -\left(\frac{36}{5}g_1^2+24 g_2^2\right) \lambda_2 +2\lambda_1^2+2\lambda_1 \lambda_4+28\lambda_2^2+24\lambda_2 \lambda_3+6\lambda_3^2 \;,\nonumber\\
	(4\pi)^2 \frac{d \lambda_3}{d t}&=&\frac{72}{5} g_1^2 g_2^2 -6  g_2^4+ \lambda_4^2 -\left(\frac{36}{5}g_1^2+24 g_2^2\right) \lambda_3 +24\lambda_2 \lambda_3+18\lambda_3^2 \;,\nonumber\\
		(4\pi)^2 \frac{d \lambda_4}{d t}&=&\frac{36}{5} g_1^2 g_2^2  -\left(\frac{9}{2}g_1^2+\frac{33}{2} g_2^2\right) \lambda_4 +\lambda \lambda_4+8\lambda_1 \lambda_4+4\lambda_4^2+4\lambda_2 \lambda_4+8\lambda_3 \lambda_4 +6\lambda_4 y_t^2 \;.
\end{eqnarray}
Here $t=\ln (\mu/M_t)$ with top mass $M_t$ and all couplings except the top case are ignored.
Based on the equations, we can obtain the evolution of quartic couplings up to the Planck mass. As shown in Ref.~\cite{Chun:2012jw}, the quartic couplings show the upward trend so that the perturbation condition is violated to cause $\lambda_i>4\pi$. However, the stability conditions are still satisfied in some certain regions as shown in Ref.~\cite{Bonilla:2015eha}. Actually, our model parameters are in this suitable region.
\\

\noindent
{\bf Yukawa couplings and Neutrino mass}

The introduction of the $Y=1$ triplet scalar will be able to generate Majorana neutrino masses via interactions with left-handed lepton doublet $L_L=(\nu_L, l_L)^T$
\begin{eqnarray}
	L_Y &=& y_{ij}\bar L_{iL}^c i \tau_2 \Delta  L_{jL}+h.c. = y_{ij} \left[
	\bar \nu_{iL}^c \Delta^0 \nu_{jL}-\bar l_{iL}^c \Delta^{++} l_{jL}-\frac{1}{\sqrt{2}}(\bar l_{iL}^c \Delta^{+} \nu_{jL}+\bar \nu_{iL}^c \Delta^{+} l_{jL})
 	\right]+h.c.\;.
\end{eqnarray}
Once the triplet acquires non-zero vev, the first term in the above equation will induce a Majorana-type mass term as
\begin{eqnarray}
	m^{ij}_\nu &=& y_{ij}\sqrt{2}v_\Delta\;.
\end{eqnarray}
Due to the neutrino mass with $O(eV)$ or less, $y_{ij}$ should have a magnitude of order $< 10^{-5}$ for $v_\Delta \sim 0.1$ GeV. 

A non-zero $y_{ij}$ will  contribute to some of the well known experimental observables such as flavor changing  decays $l_i \to l_j l_k \bar l_m$,  $l_i \to  l_j \gamma$ and lepton $g-2$. For  $y_{ij} \sim 10^{-5}$ and  $v_\Delta \sim 0.1$  GeV, we have checked numerically that  the new contributions turn out to be negligibly small to ensure compliance with current rare processes constraints as shown in \cite{Heeck:2022fvl}. 

The small  Yukawa coupling $y_{ij}$ makes almost impossible collider search for the case of new signals by $\Delta$ mediated decays. One, however, may still seek for signals through effects of $\Delta$ couplings to the gauge boson. We provide some details next.
\\

\noindent
{\bf Higgs data}

The couplings of 125-GeV Higgs  boson in the SM or the SM-like Higgs in NP has been studied extensively at the LHC. These Higgs data provide a nontrivial constraint on the type-II seesaw parameter space. These couplings are parameterized by 
\begin{eqnarray}
	L_h  &=& C_W \frac{2m_W^2}{v}h W_\mu^+ W^{-\mu}+ C_Z \frac{m_Z^2}{v}h  Z_\mu Z^{\mu}-C_f \frac{m_f}{v}h \bar f f -C_+\frac{2m_{H^+}^2}{v}h H^+ H^- 
	-C_{++}\frac{2m_{H^{++}}^2}{v}h H^{++} H^{--}\nonumber\\
   &+&\tilde C_g \frac{\alpha_s}{12\pi v}h G^{\mu\nu}G_{\mu\nu}	+
   \tilde C_\gamma \frac{\alpha}{\pi v}h F^{\mu\nu}F_{\mu\nu}	
   \;.
\end{eqnarray}
Note the couplings in the first line and the last line arise at tree level and one-loop level, respectively.  Here $C_i$'s are coupling modifiers with 
\begin{eqnarray}
	C_W  &=& \frac{\cos \alpha \;v+2\sin \alpha \;v_\Delta}{v_{SM}}\;,\;\;\; C_Z  =\frac{(\cos \alpha \;v+4\sin \alpha \;v_\Delta)v_{SM}}{v^2_{SM}+2v_\Delta^2}\;,\;\;\; C_f=\cos \alpha \frac{v_{SM}}{v}\;,\;\;\; \tilde C_g=C_f \tilde C_g^{SM}\;.
\end{eqnarray}
In the limit of small $v_\Delta$ and small $\sin \alpha$, they can reduce the SM value 
$C_f \approx C_W \approx  C_Z \approx 1$, $\tilde C_g \approx \tilde C_g^{SM} \approx 0.97$, and $C_+=(\lambda_1+\lambda_4/2)v^2_{SM}/(2m^2_{H^+})$,  $C_{++}=\lambda_1 v^2_{SM}/(2m^2_{H^{++}})$. The more complicated $\tilde C_\gamma$ can be expressed as \cite{Kanemura:2012rs,Primulando:2019evb}
\begin{eqnarray}
	\delta_\gamma  &=& \tilde C_\gamma  -\tilde C^{SM}_\gamma 
	\approx \frac{C_+}{24}A_s\left(\frac{m_h^2}{4m^2_{H^+}}\right)
	+\frac{C_{++}}{6}A_s\left(\frac{m_h^2}{4m^2_{H^{++}}}\right)
	\;,
\end{eqnarray}
with 
\begin{eqnarray}
	&&A_s (\tau)  = \frac{3}{\tau^2}[F(\tau)-\tau]\;,\;\;\;
	F(\tau)=\left\{
	\begin{array}{cl}
		\arcsin^2 \sqrt{\tau}\;,\;\; &  \;\;\; \tau \le 1 \\
	-\frac{1}{4}\left[\ln \frac{1+\sqrt{1-1/\tau}}{1-\sqrt{1-1/\tau}}-i\pi\right]^2\;,\;\; &  \;\;\; \tau > 1
	\end{array} \right.
\end{eqnarray}
Here $\tilde C^{SM}_\gamma \approx -0.81$. We find that the one-loop contributions from  doubly charged Higgs and singly charged Higgs have the positive sign, which is destructive  to the negative contribution SM prediction.

Therefore, we find that the relevant Higgs data is $h\to \gamma\gamma$ decay with signal strength via our fit results
\begin{eqnarray}
	&&R_{\gamma\gamma}   = \frac{\Gamma(h\to \gamma\gamma)}{\Gamma(h\to \gamma\gamma)^{SM}}=1-2.43 \Re(\delta \gamma)+1.5|\delta_\gamma|^2\approx 0.48\pm 0.02\;.
\end{eqnarray}
We find that the value is consistent within $2\sigma$ errors  with ATLAS~\cite{ATLAS:2018hxb} and CMS~\cite{CMS:2018piu} search channels by different production mode of the Higgs boson: gluon fusion (ggF), vector boson fusion (VBF), associated production with a vector boson (VH) and associated production with a pair of $t\bar t$ (ttH). And our value is totally agreement with the analysis in Ref.~\cite{Kanemura:2012rs}.\\

\noindent
{\bf Triple Higgs self-coupling}

The triplet scalar will affect the triple Higgs boson coupling at one-loop contributions. The leading contribution to the  coupling constant $\lambda_{hhh}$ compared to the SM prediction is obtained for $v_\Delta<<v$ as
\begin{eqnarray}
	&&k_\lambda=\frac{\lambda_{hhh} }{\lambda^{SM}_{hhh}}  \approx 1+ \frac{ 1}{12\pi^2 m_h^2 v_{SM}^2}(2m^4_{H^{++}}+2m^4_{H^{+}}+m_A^4+m_H^4)\approx 1.05\pm 0.01\;.
\end{eqnarray}
We find that $k_\lambda$ depend on the additional  scalar mass. Due to the mass splittings in Eq.~(\ref{mass splittings}), $k_\lambda$ is only determined by three parameters: $m_{H^{++}}$, $\lambda_4$ and $v_\Delta$.  Using our fit  values, we obtain that the extra triplet will enhance $k_\lambda$ around 0.05. This  is consistent with the experimental values $k_\lambda=4.6^{+3.2}_{-3.8}$~\cite{Aoki:2012yt}.  
\\

\noindent
{\bf Production and detection for doubly charged Higgs in the lepton colliders }

While a number of searches at the LHC are ongoing to experimentally verify the presence of the doubly-charged Higgs boson, the potential new scalar signals with a spectrum preferred  by CDF $m_W$ has been also studied at the LHC~\cite{Bahl:2022gqg}. The pair production cross section at the LHC is small and  the presence of numerous backgrounds further weakens its discovery prospects. 
Therefore,  a lepton collider with a much cleaner environment will be more suitable to search the mass regime of the doubly charged Higgs boson~\cite{Agrawal:2018pci}. Here we focus on the future $e^+e^-$ collider to explore the discovery prospects. 
 
We study pair production  $H^{++}H^{--}$ because $H^{++}$ is the lightest new scalar.
For a $e^+e^-$ collider, the pair production  can be obtained via virtual $\gamma^*/ Z^* $ exchange. The corresponding cross section is~\cite{Gunion:1989ci,Gunion:1996pq}
\begin{eqnarray}
	&&\sigma(s)=\frac{4\pi \alpha^2 s}{3}  \left(1-\frac{4m^2_{H^{++}}}{s}\right)^{\frac{3}{2}}\left[P_{\gamma\gamma}+P_{\gamma Z}\frac{(2s_W^2-1)\left(2s_W^2-\frac{1}{2}\right)}{2s_W^2c_W^2}
	+P_{ZZ}\frac{(-1+2s_W^2)^2\left(\left(s_W^2-\frac{1}{2}\right)^2+s_W^4\right)}{8s_W^4c_W^4}
	\right]\;.
\end{eqnarray}
with 
\begin{eqnarray}
	&&P_{\gamma\gamma}=\frac{1}{s^2}\;,\;\;\; P_{ZZ}=\frac{1}{(s-m_Z^2)^2+m_Z^2 \Gamma_Z^2}\;,\;\;\; P_{\gamma Z}=\frac{s-m_Z^2}{s} P_{ZZ}\;.
\end{eqnarray}
To generate the pair production, the center of mass energy must satisfy $\sqrt{s}\ge 2m_{H^{++}}=206$ GeV.  The future  $e^+e^-$ colliders satisfying the above condition include ILC ($\sqrt{s}=0.5$ TeV), CLIC ($\sqrt{s}=1.5$ TeV), FCC-ee ($\sqrt{s}=350$ GeV)  and  CEPC ($\sqrt{s}=360$ GeV) with the corresponding Luminosity of $2.5\times 10^{34}\;cm^{-2}s^{-1}$~\cite{Abe:2001grn},  $1\times 10^{35}\;cm^{-2}s^{-1}$~\cite{CLICPhysicsWorkingGroup:2004qvu}, $3.8\times 10^{34}\;cm^{-2}s^{-1}$~\cite{FCC:2018evy} and $0.83\times 10^{34}\;cm^{-2}s^{-1}$~\cite{CEPCPhysicsStudyGroup:2022uwl}, respectively. We obtain the cross section and event number  each year as
\begin{eqnarray}\label{event}
	&&\mbox{ILC}: \;\;\sigma(e^+e^-\to H^{++}H^{--})=366\; fb\;,\;\;\; N=9\times 10^4/year\;,\nonumber\\
		&&\mbox{CLIC}: \;\;\sigma(e^+e^-\to H^{++}H^{--})=52\; fb\;,\;\;\; N=5\times 10^4/year\;,\nonumber\\
			&&\mbox{FCC-ee}: \;\;\sigma(e^+e^-\to H^{++}H^{--})=527\; fb\;,\;\;\; N=2\times 10^5/year\;,\nonumber\\
				&&\mbox{CEPC}: \;\;\sigma(e^+e^-\to H^{++}H^{--})=520\; fb\;,\;\;\; N=4\times 10^4/year\;.
\end{eqnarray}
We find that the future lepton colliders can produce amount of event number around $O(10^4)$, which is conducive to investigate the feature of doubly charged scalars. 

To further detect the doubly charged scalars, one needs to study the decay modes. 
For the fit results in Table-\ref{table3}, the best fit is $v_\Delta=0.09$ GeV. In this case,  the dominant decay mode is two $W$'s final states as shown in Eq.~(\ref{br}). Therefore, the decay branching ratio for two final $W$ states is totally 100$\%$.  This indicates that the search for doubly charged Higgs can be conducted by involving a final state of four W bosons. LHC has conducted the dedicated search for this signature~\cite{ATLAS:2018ceg,ATLAS:2021jol}. The future lepton colliders can further explore this channel.
Combining the previous $H^{++}$ production cross section,  $\sigma(e^+ e^- \to H^{++}H^{--} \to 4W)$ could be around $O(100fb)$ and further produce $O(10^{4})$ events as shown in Eq.~(\ref{event}), which is possible to detect with enough significance.   
The $H^{++}$ mass  can be obtained via the resonance structure of $W^+W^+$ as shown in Ref.~\cite{Bahl:2022gqg}. The decay widths for other new scalars are shown in  Ref.~\cite{Aoki:2011pz, Ashanujjaman:2021txz}.
\\

\subsection{Discussion for $Y=0$ triplet model}

For $Y=0$ case,   the potential parameters can be expressed in terms of physical Higgs mass and mixing angle $\theta$ as well as $v_\Sigma$ via the help of Eqs.~(\ref{chargedmass}, \ref{neutralmass}). Using the fit values in Table-\ref{table4}, we calculate these parameters as
\begin{eqnarray}\label{unitarity}
	a_1 &=& \frac{m^2_{H^+}}{v_\Sigma(1+v^2/4v_\Sigma^2)}=10.10\pm 4.09\;,\nonumber\\
	\lambda_0 &=& \frac{m^2_h(\cos 2\theta+1)-m^2_{H^0}(\cos 2\theta-1)}{4v^2}\approx \frac{m^2_h}{2v^2}=0.13\pm 0.00002\;,\nonumber\\
	a_2 &=& \frac{a_1 v+(m^2_{H^0}-m^2_h)\sin 2\theta}{2v v_\Sigma}\approx
	\frac{a_1 }{2 v_\Sigma}=1.31\pm 0.52\;,\nonumber\\
	b_4 &= &\left[\frac{m^2_{H^0}(\cos 2\theta+1)-m^2_h(\cos 2\theta-1)}{4}  -\frac{m^2_{H^+}}{1+v^2/4v_\Sigma^2}\frac{v^2}{8v_\Sigma^2}\right]/v_\Sigma^2\approx
	\frac{\left[\frac{m^2_{H^0}}{2}  -\frac{m^2_{H^+}}{1+v^2/4v_\Sigma^2}\frac{v^2}{8v_\Sigma^2}\right]}{v_\Sigma^2}=31.27\pm 27.53
	\;.
\end{eqnarray}
Here we use the approximation $\theta \sim 0$.
Note that  $a_1$, $\lambda_0$, $a_2$ in Eq.~(\ref{unitarity})  can easily satisfy the perturbativity unitarity and vacuum stability $0<(\lambda_0, a_1, a_2)<4\pi$~\cite{Khan:2016sxm,DiLuzio:2017tfn}.  However,  the term $b_4$ in Eq.~(\ref{unitarity}) violates strongly the perturbativity unitarity.

\begin{figure}[!t]
	\centering
	\subfigure[\label{b4limit} $b_4$ allowed region]
	{\includegraphics[width=.486\textwidth]{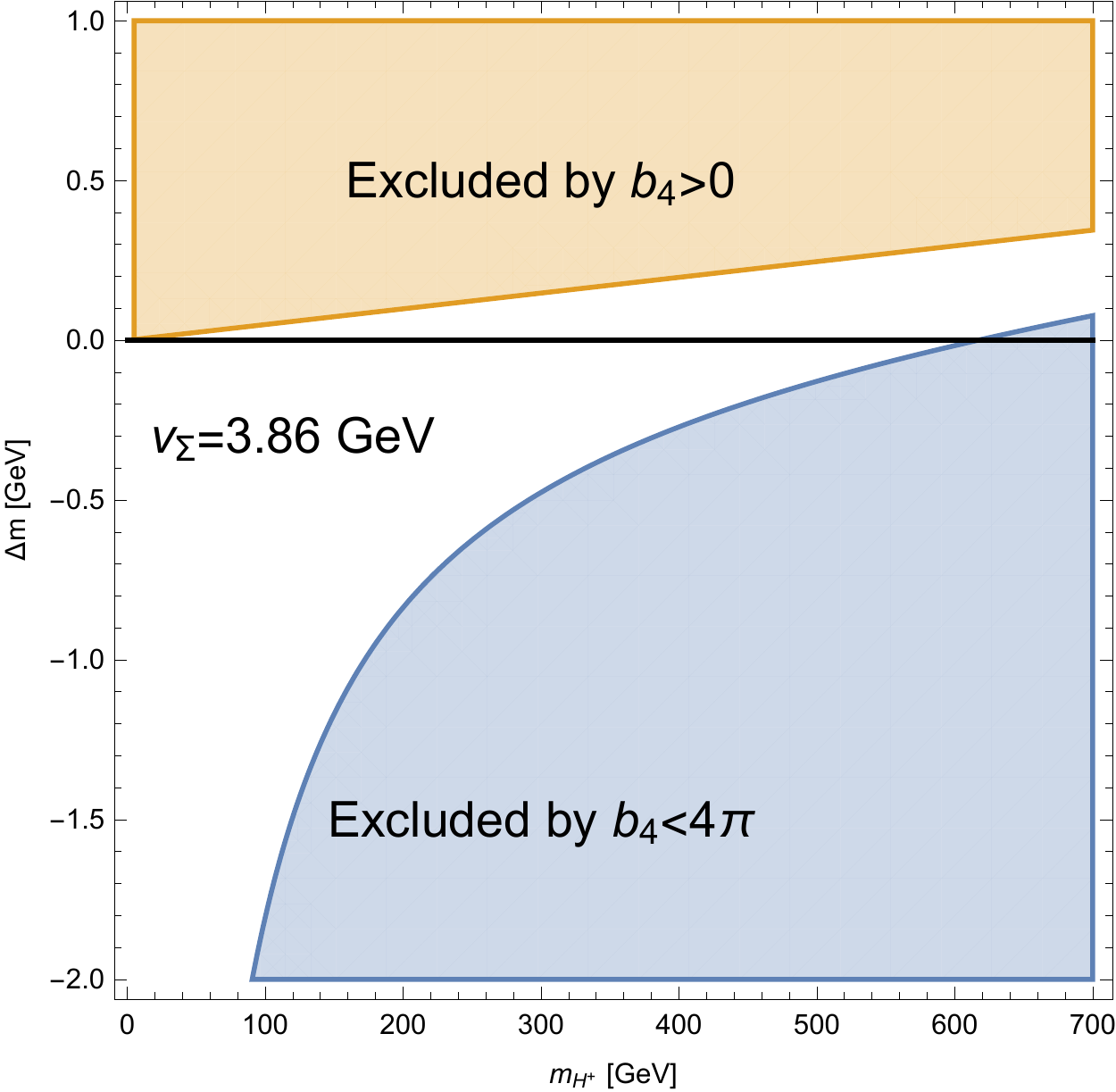}}
	\subfigure[\label{3sigma}   the fit results within $1\sigma$ errors]
	{\includegraphics[width=.486\textwidth]{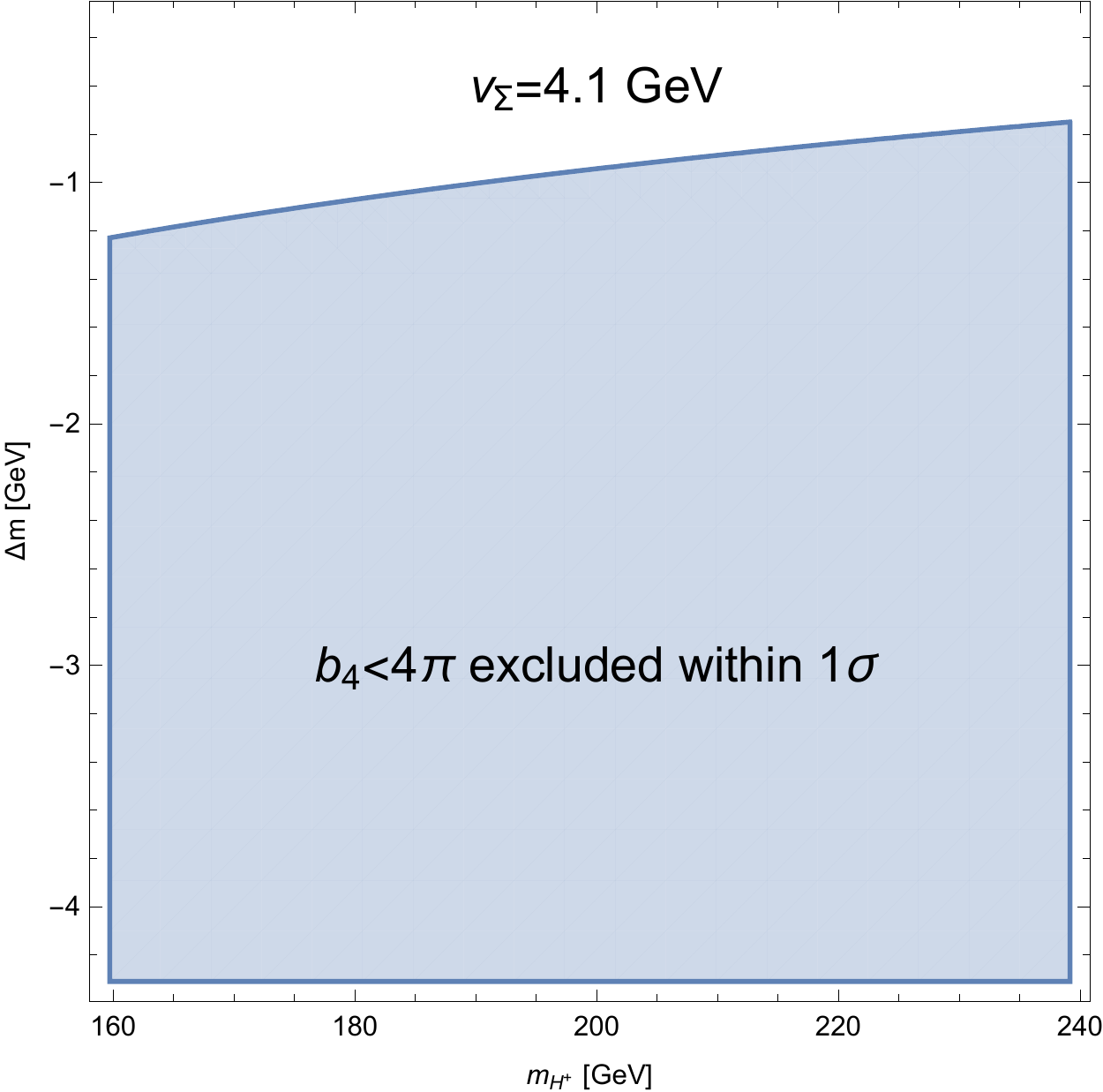}}
	\caption{The parameter $b_4$ allowed region in $\Delta m - m_{H^+}$ plane. The excluded region by vacuum stability bound $b_4>0$ is in orange, perturbativity unitraity $b_4<4\pi$ in blue.
	}
	\label{b4}
\end{figure}

Therefore, we need to consider firstly the  constraints $0<b_4<4\pi$. In the limit of $v_\Sigma<<v$, we find $b_{4}$ in Eq.~(\ref{unitarity}) can be approximated as $(m^2_{H^0}-m^2_{H^+})/(2v_\Sigma^2)$, which strongly depends on the mass difference $\Delta m= m_{H^+}-m_{H^0}$ and $v_\Sigma$. If the difference is much larger or $v_\Sigma$ is much smaller, the above constraint will be severely ruined.
Based on the $b_4$ term in Eq.~(\ref{unitarity}), we plot the  allowed region for $b_4$  with the fit value $v_{\Sigma}=3.86$  GeV as shown in Fig.~\ref{b4limit}.    
We find that the best fit values $m_{H^+}=199.45$ GeV and $\Delta m=-2.32$ GeV has been excluded by the $b_4$ constraint. And the mass difference must be much small so that the neutral and charged triplets are almost mass degenerate. 
By adopting the constraint $0<b_4<4\pi$  to  perform the global fit again, we obtain  much larger $\chi^2 \sim 35$ and  find that the flat fit values are insensitive to $m_{H^+}$.
%Under the $b_4$ constraint situation, we try to perform the global fit again. Unfortunately, we can not find the reasonable fit values with much larger $\chi^2 \sim 35$. Especially the $m_{H^+}$ fit values are flat depending  on the specific ranges, which shows the insensitive to the mass $m_{H^+}$.  
Because the $b_4$ parameter has the stringent dependence on $\Delta m$, we try to decrease the size of  $\Delta m$ by considering the error bars. 
Based on the number $b_4=31.27\pm 27.53$, we find that  the $b_4$ constraints can  be satisfied within $1\sigma$ where  $\Delta m$ can decrease to -0.33 GeV. In this case, we plot the $b_4$ allowed region for the fit results within $1\sigma$ as shown in Fig.~\ref{3sigma}.  Therefore, we find that the fit results within $1\sigma$ errors can satisfy the $b_4$ constraint.

Here  we  give a simple phenomenological analysis. Due to $m_{H_1}=m_h=125$ GeV  and $\theta \to 0$, $H^0$ will do not couple to SM fermions nor to the Z gauge boson. 
Therefore, $H^0$ is almost  fermiophobic so that  the dominant two-body decay is $H^0\to WW$. In the case of charged Higgs, the main decay modes are $H^+ \to t\bar b$, $H^+\to W^+ Z$ and $H^+\to W^+h$, as shown in Ref.~\cite{FileviezPerez:2022lxp}. The collider phenomenology of $Y=0$ triplet has been also studied in Refs.~\cite{FileviezPerez:2008bj,Chiang:2020rcv}.

\section{Conclusion}

Electroweak precision observables are fundamental for testing the SM or its extensions. The influences to observables from new physics  within the electroweak sector can be  expressed in terms of oblique parameters $S$, $T$, $U$.  By performing the global fit with the CDF new $W$ mass, we obtain $S=0.03 \pm 0.03$, $T=0.06 \pm 0.02$ and $U=0.16 \pm 0.03$ or $S=0.14 \pm 0.03$, $T=0.24 \pm 0.02$ with $U=0$, which strongly indicate the need of  the new physics beyond SM. 
We have studied $Y=1$ and $Y=0$ triplet models influences on 
the oblique parameters and other observables. In both models, there are tree and loop level contributions to the $S$, $T$ and $U$ parameters simultaneously. We carry out global fits by  the triplet model parameters directly trying to find the allowed parameter space of these two models.

For $Y=1$ triplet model,  the tree contribution due to a non-zero vev $v_\Delta$ has the wrong sign to solve the CDF new $W$ mass excess problem. However,  non-degenerate scalars in the model at loop level contribute to address the $W$ mass excess problem. In this model there are seven physical scalars, doubly charged $H^{\pm\pm}$, singly charged $H^{\pm}$, neutral CP-even $H$, SM-like $h$, and neutral CP-odd $A$. 
Their effects can be expressed in terms of three parameters, the doubly charged  mass $m_{H^{++}}$, potential parameter $\lambda_4$ and triplet vev $v_\Delta$. Then we perform the global fit  to obtain $m_{H^{++}}=103.02 \pm 9.84$ GeV, $\lambda_4=1.16\pm 0.07$  and $v_\Delta=0.09 \pm 0.09$ GeV.
The corresponding best value for $S$, $T$ and $U$ are  $S=0.15$, $T=0.22$, $U=0.04$, which are consistent with by treating $S$, $T$ and $U=0$ independently.
 These fit values can result in mass difference  $\Delta m=m_{H^+}-m_{H^{++}}=64.78\pm 2.39$ GeV and doubly charged mass $m_{H^{++}}$ satisfies the current collider constraints.
 Furthermore, we also analyze the relevant phenomenology, such as  the vacuum stability, perturbative unitarity, Higgs data, triple Higgs self-coupling and lepton colliders analysis.
%And we find that the future  can feasibly search for doubly charged $H^{++}$    via pair production $H^{++}H^{--}$ and decay  channel to four W final states.

For $Y=0$ triplet model, the tree level contribution helps to easy the $W$ mass excess problem. At loop level the four physical scalars,  singly charged $H^{\pm}$, neutral CP-even $H$ and SM-like $h$ will additionally affect the $W$ mass. 
Their effects can be parameterized  by the singly charged mass $m_{H^+}$, the mass difference 
$\Delta m=m_{H^+}-m_{H^0}$ and triplet vev $v_\Sigma$. 
By performing the global fit, we obtain the $m_{H^+}=199.45 \pm 39.73$ GeV, 
$\Delta m=-2.32\pm1.99$ GeV and $v_\Sigma=3.86\pm 0.27$ GeV. Under  this situation,  the best fit values of the model parameter give $S=-0.003$, $T=0.13$, $U=0.002$. Note that  $U$ is not much smaller than $S$, which clearly is contradictory to the expectation that $U$ is negligible compared with $S$ and $T$ in the NP models. More importantly, these strongly violates the perturbative  unitarity of the  potential parameter $b_4$, which can only be satisfied within $1\sigma$ errors.  %Therefore, we find that the global fit results can be compatible with the parameter $b_4$ perturbative unitarity within $1\sigma$ errors.  

%{\color{blue}The following may not needed. }
%The EW fit influences on NP are usually shown in terms of the oblique parameters. However, for a given  specific NP models, the correlations between $S$, $T$ and $U$ are different. Therefore, a global fit stopping at these oblique parameters may not provide intricate detail about the models and we need to perform the fit via directly NP parameters. We first study  the two kinds of triplet models to conduct the fits by the corresponding new parameters. This provides much explicit details for the triplet models and our analysis can further be extended to other NP models. }  

\begin{acknowledgments}
	
We thank Julian Heeck for very nice suggestions and Prof. Yong-Cheng Wu  for useful discussions.	This work was supported in part by the NSFC (Grant Nos. 11735010, 11975149, and 12090064). XGH was supported in part by the MOST (Grant No. MOST 109-2112-M-002-017-MY3). ZPX was supported  by the NSFC (Nos. 12147147).\\
\end{acknowledgments}

\end{document}